\documentclass[journal]{IEEEtran}
\usepackage{cite}
\usepackage{amsmath,amssymb,amsfonts}
\usepackage{graphicx}
\usepackage{textcomp}
\usepackage{xcolor}

\usepackage{hyperref}
\usepackage{amsmath}
\usepackage{autobreak}
\usepackage{amsfonts}
\usepackage{bm}
\usepackage{bbm}
\usepackage{algorithm}  
\usepackage{algorithmicx}  
\usepackage{algpseudocode}
\usepackage{stfloats}
\usepackage{cuted}
\usepackage{color}
\usepackage{xcolor}
\usepackage{subfigure}
\usepackage{cite}
\usepackage{booktabs} 
\usepackage{soul}
\definecolor{red}{RGB}{0,0,0}
\newcommand{\sto}{\text{s.t.}}
\newcommand{\diag}{\text{diag}}
\newcommand{\tr}{\text{tr}}
\newcommand{\vect}{\text{vec}}
\newcommand{\vxi}{\tilde{\bm\xi}}
\newcommand{\vvarphi}{\tilde{\bm\varphi}}
\newcommand{\valpha}{\tilde{\bm\alpha}}
\newcommand{\crb}{\text{CRB}}
\newcommand{\Frf}{\mathbf F_{RF}}
\newcommand{\Fbb}{\mathbf F_{BB}}
\newcommand{\Rth}{R_{th}}
\newcommand{\Rx}{\text{R}_\mathbf{x}}
\newcommand{\red}[1]{{\color{red}#1}}
\newcommand{\JppU}{\mathbf J_{\vvarphi\vvarphi}(\mathbf U)}
\newcommand{\JpaU}{\mathbf J_{\vvarphi\valpha}(\mathbf U)}
\newcommand{\JaaU}{\mathbf J_{\valpha\valpha}(\mathbf U)}
\newcommand{\grad}{\text{grad}}

\newcommand{\bvpi}{\bm\varpi}

\ifCLASSINFOpdf

\else

\fi

\begin{document}
	
	\title{CRB Minimization for RIS-aided mmWave Integrated Sensing and Communications}
	%
	%
	%
	
	\author{Wanting Lyu,~Songjie Yang,~Yue Xiu,~Ya Li,~Hongjun He, \\ ~Chau Yuen,~\IEEEmembership{Fellow,~IEEE},~and~Zhongpei Zhang,~\IEEEmembership{Member,~IEEE} 
    
		\thanks{Wanting Lyu, Songjie Yang, and Yue Xiu are with National Key Laboratory of Wireless Communications, University of Electronic Science and Technology of China, Chengdu 611731, China (E-mail: lyuwanting@yeah.net; yangsongjie@std.uestc.edu.cn; xiuyue12345678@163.com).

        Ya Li and Hongjun He are with Future Research Laboratory, China Mobile Research Institute, Beijing, China (E-mail: liyayjy@chinamobile.com; hehongjun@chinamobile.com).

        Chau Yuen is with the School of Electrical and Electronics Engineering, Nanyang Technological University, 639798 Singapore (E-mail: chau.yuen@ntu.edu.sg).
        
        Zhongpei Zhang is with the Shenzhen institute for Advanced Study, University of Electronic Science and Technology of China, Shenzhen 518110, China(e-mail:zhangzp@uestc.edu.cn).}}

	\maketitle
	
	\begin{abstract}
		In this paper, reconfigurable intelligent surface (RIS) is employed in a millimeter wave (mmWave) integrated sensing and communications (ISAC) system. To alleviate the multi-hop attenuation, the semi-self sensing RIS approach is adopted, wherein sensors are configured at the RIS to receive the radar echo signal. Focusing on the estimation accuracy, the Cram$\acute{\text{e}}$r-Rao bound (CRB) for estimating the direction-of-the-angles is derived as the metric for sensing performance. A joint optimization problem on hybrid beamforming and RIS phase shifts is proposed to minimize the CRB, while maintaining satisfactory communication performance evaluated by the achievable data rate. The CRB minimization problem is first transformed as a more tractable form based on Fisher information matrix (FIM). To solve the complex non-convex problem, a double layer loop algorithm is proposed based on penalty concave-convex procedure (penalty-CCCP) and block coordinate descent (BCD) method with two sub-problems. Successive convex approximation (SCA) algorithm and second order cone (SOC) constraints are employed to tackle the non-convexity in the hybrid beamforming optimization. To optimize the unit modulus constrained analog beamforming and phase shifts, manifold optimization (MO) is adopted. Finally, the numerical results verify the effectiveness of the proposed CRB minimization algorithm, and show the performance improvement compared with other baselines. Additionally, the proposed hybrid beamforming algorithm can achieve approximately 96\% of the sensing performance exhibited by the full digital approach within only a limited number of radio frequency (RF) chains.
	\end{abstract}
	
	\begin{IEEEkeywords}
		Integrated sensing and communications, reconfigurable intelligent surface, millimeter wave, Cram$\acute{\text{e}}$r-Rao bound, beamforming.
	\end{IEEEkeywords}
	
	\IEEEpeerreviewmaketitle
	
	\section{Introduction}

	The integration of sensing capabilities into wireless communication networks has emerged as a prominent feature in the development of beyond fifth generation (B5G) and sixth-generation (6G) mobile networks \cite{Fanliu_JSAC}. To reduce the hardware cost and increase the spectral- and energy-efficiency, sharing the same hardware platform and time-frequency resources between radar and communications has been attractive for both academia and industrial \cite{8892631}. Moreover, integrated sensing and communications (ISAC) provides the opportunity for wireless networks to collect the sensory data to establish the smart radio environment from the surrounding areas, boosting the development of environment-aware technologies \cite{9652071}. For example, in various application scenarios such as autonomous driving, industrial internet of things, environmental monitoring and augmented reality, the acquisition of precise sensory data and real-time transmission are of paramount importance. These fundamental requirements significantly contribute to enhancing safety measures, optimizing industrial workflows, augmenting user experiences, and other critical aspects. Hence, enhancing the performance of the ISAC system design has significant importance in the successful implementation of B5G and 6G applications.
	
	Millimeter-wave (mmWave) has been widely explored in wireless communications and sensing networks to provide large bandwidth, high directional beams, and high sensing resolution. Nevertheless, the practical implementation of mmWave ISAC faces considerable challenges attributed to severe blockage and path loss resulting from the short wavelength characteristics of mmWave signals \cite{7160780}. Although multiple-input multiple-output (MIMO) technologies can be adopted to mitigate the signal attenuation with multiplexing gain and beamforming gain, mmWave signals remain susceptible to obstruction by various obstacles including buildings, trees, and moving human bodies. To address this issue, reconfigurable intelligent surface (RIS) has become a promising technology to enhance wireless signal and provide extra reflecting links. Configuring with a large number of reflecting elements, RIS is able to adjust the phase of the incident signal to reconfigure the radio environment \cite{8811733,10083178}. Meanwhile, RIS-aided systems can achieve high cost- and energy- efficiency because the reflecting elements are passive, thereby eliminating the need for decoding or radio frequency (RF) processing. Based on the above advantages, this paper considers a RIS-assisted mmWave ISAC network to achieve high sensing accuracy while guaranteeing the communication performance.
	
	\subsection{Related Works}
	
	Recent years have witnessed the fast development of ISAC system design. To enhance both the illumination power at the target and the receive signal-to-interference-plus-noise (SINR) at the communication users, joint beamforming design for radar and communication signals are processed at the base station (BS) \cite{8288677,9124713,10086626,9668964,Peng,Lyu}. In \cite{8288677}, separated and shared deployment for dual-functional radar-communication (DFRC) BS were investigated respectively, where in the former, antennas were divided into two groups transmitting sensing and communication signals separately; while in the later, a joint waveform was shared by both functions. In \cite{9124713}, the authors studied joint transmit beamforming design of a dual-functional multi-user and MIMO radar system, where the mean square beampattern matching error was minimized with the constraints on the minimum received SINR of the communication users. In addition to the conventional beampattern matching as the sensing performance criterion, authors in \cite{10086626} explored the maximization of the weighted minimum beampattern gain, which exhibited superior performance in terms of worst-case beampattern gain and computational complexity. To cater for the ever-increasing number of accesses, non-orthogonal multiple access (NOMA) was applied in ISAC, where authors in \cite{9668964} investigated the beamforming design problem in NOMA-ISAC network. To further involve the estimation accuracy into sensing performance evaluation, Cram$\acute{\text{e}}$r-Rao bound (CRB), which describes the lower bound of the variance of an unbiased estimator, has been employed as a metric in ISAC \cite{9652071,Yang1,RISsensing,9591331}. In \cite{9652071}, CRBs for both point-like target and extended target cases were derived and minimized as the objective functions respectively. This was achieved through optimizing the transmit beamforming while satisfying the communication SINR threshold and adhering to the transmit power budget by optimizing the transmit beamforming, satisfying communication SINR threshold within the transmit power budget.
	
	To enhance the beamforming gain and provide extra degrees of freedom in wireless environment, RIS has been widely investigated in mmWave communications \cite{9032163, MO_3methods,9472958}. In \cite{9032163}, the mean-square-error (MSE) minimization between the received and the transmitted symbols was studied by jointly optimizing the hybrid beamforming and the RIS phase shifts. Authors in \cite{MO_3methods} proposed a penalty-based optimization algorithm and a low complexity sequential optimization algorithm to solve the power minimization problem in a RIS-aided mmWave MIMO system. In \cite{9472958}, the weighted sum rate of a RIS-aided mmWave NOMA system was maximized by joint power allocation, hybrid beamforming and RIS coefficients design. To improve both sensing and communication performance, \red{some recent works have applied RIS into ISAC \cite{9729741,9844707,9769997,add1,add2,add3}}. Specifically, in \cite{9729741}, a double RIS-assisted communication radar coexistence system was exploited, where a penalty dual decomposition (PDD)-based algorithm as well as a low complexity algorithm were proposed to maximize the communication SINR while keeping radar performance. To jointly optimize the communication and sensing performance, authors in \cite{9844707} proposed a weighted sum of data rate and radar mutual information maximization problem by optimizing active and passive beamforming. In \cite{9769997}, an alternative direction method of multipliers-majorization-minimization based method was proposed to solve the RIS-aided ISAC problem. \red{In \cite{add1}, both single-RIS and dual-RIS scenarios were studied. The single-RIS was dedicated for communication function, while in dual-RIS scenario, each RIS was separately utilized to assist radar and communication, respectively. In \cite{add2}, physical layer security was investigated in RIS-aided ISAC to suppress the eavesdropper's SINR while enhancing the user's data rate. Considering target location estimation error, authors in \cite{add3} investigated a robust beamforming problem for ISAC via hybrid active-passive RIS. The worst-case sensing beampattern gain was adopted as the objective of the optimization problem.}
	
	Some recent works have considered CRB as sensing metric in RIS-assisted ISAC to better cater for the estimation accuracy requirement \cite{RISsensing,9591331}. In \cite{RISsensing}, CRBs for estimating the angle of point target and the response matrix of extended target were derived and minimized. In the simulation results, estimation MSE was derived and computed by maximum likelihood (ML) estimation to verify the effectiveness of CRB derivation and minimization design. In \cite{9591331}, RIS was utilized to mitigate the multi-user interference under the CRB constraint for target angle estimation. In \cite{liu2023snrcrbconstrained}, both signal-to-noise-ratio and CRB were studied to measure the target detection and parameter estimation performance for RIS-ISAC networks. Two sum rate minimization problems were proposed and solved based on the two sensing performance constraints, respectively.  However, target sensing with RIS could be challenging due to the multi-hop reflection path (i.e. BS $\to$ RIS $\to$ target $\to$ RIS $\to$ BS), where multiplicative fading will be introduced, resulting in severe signal attenuation. To alleviate the large fading effect, authors in \cite{9724202} proposed a RIS-self-sensing structure, where RIS is configured with both reflecting elements and sensors, and a RIS controller to send signals for sensing. Following this, in \cite{10050406}, a simultaneously transmitting and reflecting surface configured with sensing elements was applied in ISAC to minimize the CRB for estimating the angle of target. 
	
	Although the existing works have employed CRB as the optimization objective to improve the estimation accuracy in RIS-aided ISAC network,  they have solely focused on the sub-6GHz band and full digital beamforming at the BS. However, mmWave has emerged as a crucial enabler in B5G and 6G, offering substantial bandwidth for communications and delivering high-resolution capabilities for target sensing purposes. Inspired by the sensing RIS in \cite{9724202}, we consider a mmWave ISAC system aided by a RIS configured with both reflecting and sensing elements, where CRB for estimating the azimuth and elevation angles of target is derived and minimized. Different from \cite{9724202}, where the signal is transmitted by the RIS controller, the dual-functional BS is utilized to transmit both communication and sensing symbols to avoid extra cost on RIS controller. In addition, hybrid beamforming is adopted instead of full digital beamforming for mmWave system to reduce the hardware cost.
	
	\subsection{Main Contributions}
	
	In this paper, we explore sensing-RIS into mmWave ISAC to serve multiple users and one single target. The main contributions of this paper are summarized as follows.
	
	\begin{itemize}
		\item We consider a 3-dimensional (3D) downlink mmWave ISAC system with a sensing-RIS. Specifically, the RIS is configured with both reflecting and sensing elements. The transmitted signal experiences two hops for sensing, i.e. BS $\to$ RIS reflecting elements $\to$ target $\to$ RIS sensing elements. Also, the RIS is used to serve the communication users by providing reflecting paths. 
	    
	    \item We derive the CRB for estimating the angles of target by taking the estimation accuracy into consideration for evaluating sensing performance. The communication performance is measured by the data rate. Thus, we propose a CRB minimization problem while ensuring the communication data rate for each user by jointly optimizing the hybrid beamforming matrices and the RIS coefficient matrix.
	    
	    \item To solve the highly coupled non-convex problem, we first reformulate it by introducing auxiliary variables and Schur complement. To deal with the equality constraints, penalty terms are introduced to the new objective function. For the coupled variables, we decompose the problem into two sub-problems based on block coordinate descent (BCD) method, and iteratively optimize the variables in two groups. 
	    
	    \item For optimizing the hybrid beamforming as well as the digital beamforming matrices, we propose a penalty concave-convex procedure (penalty-CCCP) based algorithm, with successive convex approximation (SCA) method and second order cone (SOC) constraints. For optimizing the analog beamformer as well the phase shifts matrix with unit modulus constraints, a complex circle manifold (CCM) based optimization algorithm is proposed with two independent Riemannian manifolds.
	    
	    \item The numerical results show the effectiveness of the proposed algorithm and the performance improvement in terms of sensing CRB, compared with other baselines. The trade-off between sensing and communication performance is analyzed as well. To verify the CRB minimization method, we further compute the MSE of the target angle estimation with the optimal beamforming and phase shifts design. The results show that the proposed algorithm is effective in improving the estimation accuracy through the reduction of MSE.
	\end{itemize}
	
	\subsection{Organizations and Notations}

    The rest of the paper is organized as follows. In Section II, we give the system model of the RIS aided mmWave ISAC network, and derive the metrics for communication and sensing, respectively. In Section III, the optimization problem is proposed to minimize the CRB for angle estimation, with constraints on communication data rate. A BCD-based algorithm is proposed to solve the complex non-convex problem. Section IV summarizes the overall algorithm, and analyzes the convergence and computational complexity. Thu numerical results are illustrated and presentated in Section V, and the overall contents of this paper is concluded in Section VI.
 
    In this paper, vectors and matrices are denoted by boldface lower and upper-case letters, respectively. Conjugate, transpose and conjugate transpose of $\mathbf x$ are denoted by $\mathbf x^*$, $\mathbf x^T$ and $\mathbf x^H$, respectively. $||\mathbf x||$ denote the $l_2$ norm of vector $x$, and $||\mathbf X||_F$ is the Frobenius norm of matrix $\mathbf X$. $\otimes$ and $\odot$ denote the Kronecker and Hadamard product. $\mathbb E\{\cdot\}$ is the expectation operator, $\mathcal Re\{\cdot\}$ and $\mathcal Im\{\cdot\}$ are the real and imaginary part of the parameter, respectively. $\tr(\cdot)$ is the trace of the parameter, and $\vect{\cdot}$ means to verctorize a matrix. $\mathbf I_N$ is the $N\times N$ identity matrix, and $\{\cdot\}^{-1}$ is the matrix inverse operator.   
	\section{System model}
	
	\subsection{Signal Model}
	\begin{figure}[t]
		\centering
		\includegraphics[width=0.8\linewidth]{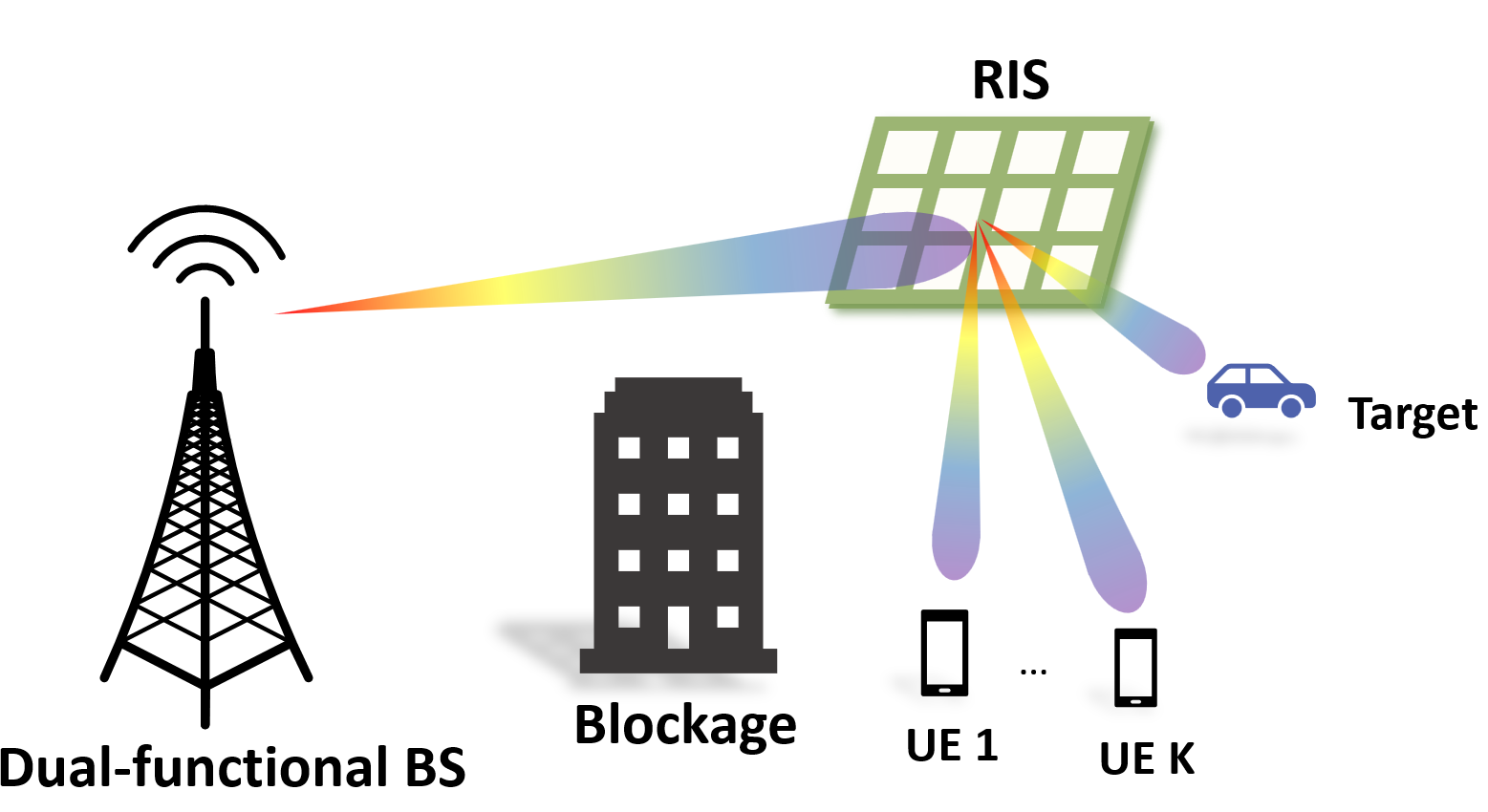}
		\caption{System model of the mmWave ISAC network.}
		\label{system model}
	\end{figure}
	Consider a downlink ISAC model working in mmWave band with a DFRC BS. The BS transmit communication signals to serve $K$ single antenna communication users and probing signals to sense a single target as illustrated in Fig \ref{system model}. Assume that the BS is configured with a uniform planar array (UPA) at the x-z plane, with $N_t = N_{tx}N_{tz}$ transmit antenna elements. To reduce the hardware power consumption, hybrid beamforming is adopted at the BS with $N_{RF}\ll N_t$ RF chains. $N_s$ data streams are generated and processed by a hybrid beamformer, which can be written as
	\begin{equation}
		\mathbf x(t) = \mathbf F_{RF}\mathbf F_{BB}\mathbf s(t),
		\label{TxSig}
	\end{equation}
	where $\mathbf F_{RF} \in \mathbb C^{N_t\times N_{RF}}$ denotes the analog beamformer, $\mathbf F_{BB} \in \mathbb C^{N_{RF}\times N_s}$ denotes the digital beamformer, and $\mathbf s(t) \in \mathbb C^{N_s\times 1}$ is the transmitted symbol. To allocate the resources for both sensing and communication functions, the transmit data symbol is defined as
	\begin{equation}
		\mathbf s(t) = \big[ \underbrace{s_1(t), \cdots,  s_K(t)}_\text{Communication}, \underbrace{s_{K+1}(t), \cdots, s_{N_s}(t)}_\text{Sensing}  \big]^T,
	\end{equation}
	consisting $K$ data streams for serving users, and $N_s-K$ streams for sensing targets. \red{Note that the combined communication symbol and sensing waveform $\mathbf s(t)$ is assumed to satisfy mutually independence as $\mathbb E\{\mathbf s(t)\mathbf s^H(t) \} = \mathbf I_{N_s}$ \cite{9124713}. The specific design of the waveform is beyond the scope of this paper.} Accordingly, the digital beamformer $\Fbb$ is defined as
	\begin{equation}
		\mathbf F_{BB} = \big[ \underbrace{\mathbf f_1, \cdots, \mathbf f_K}_\text{Communication}, \underbrace{\mathbf f_{K+1}, \cdots, \mathbf f_{N_s}}_\text{Sensing}  \big],
	\end{equation}
	We further denote the hybrid beamforming matrix $\mathbf W = \mathbf F_{RF}\mathbf F_{BB} = [\mathbf w_1, \cdots, \mathbf w_{N_s}]$. The covariance matrix of the transmit signal $\mathbf x(t)$ can be denoted as
	\begin{equation}
		\Rx = \mathbb E\{\mathbf x(t)\mathbf x^H(t)\} = \mathbf W\mathbf W^H. \label{Cov}
	\end{equation}
	
	We consider the sensing-RIS, which is also a UPA consisting of $M = M_yM_z$ reflecting elements and $M_s = M_{sy}M_{sz}$ sensing elements deployed in the y-z plane. Consider that the reflecting elements can only adjust the phases, and reflecting coefficient matrix can be represented as $\mathbf \Theta = \diag\{e^{j\theta_1},\dots,e^{j\theta_{M}}\}$, where $\theta_m\in [0,2\pi), \forall m\in\{1,\dots,M\}$.
	
	\subsection{Metrics for Communications}
	The channels from the BS to the RIS, and from the RIS to user $k$ are denoted as $\mathbf H \in \mathbb C^{M\times N_t}$, $\mathbf g_r^H \in \mathbb C^{1\times M}$, which are assumed to be perfectly known by the BS with sophisticated channel estimation methods \cite{9732214,10093915,10081022}. Without loss of generality, the direct links from the BS to the users and the target are assumed to be severely blocked by the densely distributed obstacles, which is a typical scenario in RIS aided systems \cite{9326394}.  Consequently, the received signal of user $k$ can be expressed as 
	\begin{equation}
		\begin{split}
			y_k(t) = &\; \mathbf g_k^H\mathbf\Theta\mathbf H\mathbf x(t) + n_k, \\
			= &\; \underbrace{\mathbf g_k^H\mathbf\Theta\mathbf H\mathbf w_k s_k(t)}_\text{desired signal} + \underbrace{\sum_{j=1,j\neq k}^{K}\mathbf g_k^H\mathbf\Theta\mathbf H \mathbf w_j s_j(t)}_\text{inter-user interference}  \\
            &+  \underbrace{\sum_{j=K+1}^{N_{RF}}\mathbf g_k^H\mathbf\Theta\mathbf H \mathbf w_j s_j(t)}_\text{sensing interference} + \underbrace{n_k}_\text{AWGN},
		\end{split}
	\end{equation}
	where $n_k \sim \mathcal{CN}(0,\sigma_0^2)$ denotes the additive white Gaussian noise (AWGN). The received SINR of user $k$ is 
	\begin{equation}
		\gamma_k = \frac{\left|\mathbf g_k^H\mathbf\Theta\mathbf H\mathbf w_k\right|^2}{\sum_{j=1,j\neq k}^{K}\left|\mathbf g_k^H\mathbf\Theta\mathbf H\mathbf w_j\right|^2  + \sum_{j=K+1}^{N_{RF}}\left|\mathbf g_k^H\mathbf\Theta\mathbf H\mathbf w_j\right|^2+ \sigma_0^2}.
	\end{equation}
	Thus, the achievable data rate of user $k$ is
	\begin{equation}
		r_k = \log_2(1+\gamma_k).
	\end{equation}

	\subsection{Metrics for Sensing}
	As for sensing, the target in this paper is assumed to be in the far field of the BS and the RIS, and can be viewed as a point-like target. We first define the steering vectors of RIS sensing elements and reflecting elements as
	\begin{equation}
    \begin{split}
		\mathbf a(\eta,\phi) = \frac{1}{\sqrt{M_s}}& \left[1, \cdots, e^{-j\frac{2\pi}{\lambda}(M_{sy}-1)\sin\eta\sin\phi d_y} \right]^T  \\
        \otimes &\left[1, \cdots, e^{-j\frac{2\pi}{\lambda}(M_{sz}-1)\cos\phi d_z} \right]^T, 
    \end{split}
    \end{equation}
    \begin{equation}
    \begin{split}
		\mathbf b(\eta,\phi) = \frac{1}{\sqrt{M}} &\left[1, \cdots, e^{-j\frac{2\pi}{\lambda}(M_{y}-1)\sin\eta\sin\phi d_y} \right]^T  \\
         \otimes& \left[1, \cdots, e^{-j\frac{2\pi}{\lambda}(M_{z}-1)\cos\phi d_z} \right]^T, 
    \end{split}
	\end{equation}
	where $\lambda$ denotes the carrier wavelength, $d_y$ and $d_z$ denotes the horizontal and vertical adjacent element space, respectively. $\eta$ and $\phi$ are the azimuth and elevation angles.
	
	Accordingly, the received echo signal at the sensors on RIS can be modeled as \footnote{Note that interference of the signal directly from the BS to RIS sensors can be removed, because the transmitted signal and the channel state information is known by the RIS controller.}
	\begin{equation}
		\mathbf Y_r = \alpha \mathbf a(\eta,\phi)\mathbf b^T(\eta,\phi)\mathbf{\Theta HX} + \mathbf N_r,
	\end{equation}
	where $\alpha$ denotes a complex constant containing the round-trip path loss and the radar cross section (RCS) at the target, $\mathbf a(\eta,\phi) \in \mathbb C^{M_s\times 1}$ denotes the steering vector of the RIS sensing elements, $\mathbf b^T(\eta,\phi) \in \mathbb C^{1\times M}$ denotes the steering vector the RIS reflecting elements, and $\mathbf N_r$ denotes the AWGN with each entry being $\sigma_r^2$. Consider the Swerling-I model \cite{1987Principles}, where the fluctuations of the RCS is assumed to be slow and the round-trip sensing channel is assumed to be constant during the transmission of $T$ communication and sensing symbols. Also note that the transmitted signal, received signal and the AWGN are stacked as $\mathbf Y_r = \big[\mathbf y_r(1), \cdots, \mathbf y_r(T) \big]$, $\mathbf X = \big[\mathbf x(1), \cdots, \mathbf x(T) \big]$ and $\mathbf N_r = \big[\mathbf n_r(1), \cdots, \mathbf n_r(T) \big]$, respectively, where $T$ represents the radar dwell time. When $T$ is sufficiently large, the covariance matrix in (\ref{Cov}) can be then practically derived as \cite{arxiv_STARISAC}
	\begin{equation}
		\Rx = \mathbf{WW}^H \approx \frac{1}{T}\mathbf {XX}^H.
	\end{equation}
	
	According to \cite{RISsensing}, we vectorize $\mathbf Y_r$ such that 
	\begin{equation}
		\mathbf y_r = \text{vec}(\mathbf Y_r) = \mathbf p + \mathbf n_r, 
		\label{Rx_vec}
	\end{equation}
	where $\mathbf p = \text{vec}\left( \alpha \mathbf a(\eta,\phi)\mathbf b^T(\eta,\phi)\mathbf{\Theta HX} \right)$ and $\mathbf n_r = \text{vec}(\mathbf N_r) \in \mathcal{CN}(\mathbf 0, \mathbf R_{n_r})$ with $\mathbf R_{n_r} = \sigma_{r}^2\mathbf I_{M_sT}$.

	Denote the estimated parameters for target sensing $\tilde{\bm\xi} = [\tilde{\bm\varphi}^T, \tilde{\bm\alpha}^T]^T$, where $\tilde{\bm\varphi} = [\eta, \phi]^T$ and $\tilde{\bm\alpha} = [\mathcal{R}e(\alpha), \mathcal{I}m(\alpha)]^T$. Based on \cite{1994Fundamentals}, the FIM for estimating the parameter $\tilde{\bm\xi}$ can be partitioned as
	\begin{equation}
		\mathbf J = \begin{bmatrix}
			\mathbf J_{\tilde{\bm\varphi}\tilde{\bm\varphi}}   & \mathbf J_{\tilde{\bm\varphi}\tilde{\bm\alpha}}  \\
			\mathbf J_{\tilde{\bm\varphi}\tilde{\bm\alpha}}^T   & \mathbf J_{\tilde{\bm\alpha}\tilde{\bm\alpha}}  \\
		\end{bmatrix}, \label{FIM}
	\end{equation}
	with each element of $\mathbf J$ given by
	\begin{equation}
    \begin{split}
		\mathbf J_{i,j} &= \tr\left( \mathbf R_{n_r}^{-1}\frac{\partial\mathbf R_{n_r}}{\partial\vxi_i}\mathbf R_{n_r}^{-1}\frac{\partial\mathbf R_{n_r}}{\partial\vxi_j} \right) + 2\mathcal Re\left\{ \frac{\partial\mathbf p^H}{\partial\vxi_i}\mathbf R_{n_r}^{-1}\frac{\partial\mathbf p}{\partial\vxi_j} \right\} \\
        &\overset{(a)}{=} \frac{2}{\sigma_r^2}\mathcal Re\left\{ \frac{\partial\mathbf p^H}{\partial\vxi_i}\frac{\partial\mathbf p}{\partial\vxi_j} \right\},
		\label{FIM_elem}
    \end{split}
	\end{equation}
	where (a) is because $\mathbf R_{n_r}$ is independent of $\vxi$ such that $\frac{\partial\mathbf R_{n_r}}{\partial\vxi_i} = 0$ for any $i$. The detailed derivation of the FIM elements can be seen in \textbf{Appendix A}. The CRB matrix for estimating $\vvarphi$ is given by \cite{1703855} as
	\begin{equation}
		\crb(\vvarphi) = [\mathbf J_{\vvarphi\vvarphi} - \mathbf J_{\vvarphi\valpha}\mathbf J_{\valpha\valpha}^{-1}\mathbf J_{\vvarphi\valpha}^T]^{-1}.
	\end{equation}
	In CRB matrix, each diagonal element represents the lower bound of the variance of an unbiased estimator for estimating the corresponding parameter. Thus, we minimize the trace of the CRB matrix as the objective function to improve the sensing performance. 
	
	Note that $\crb(\vvarphi)$ is related to the target angle $\vvarphi$, and minimizing the CRB is through optimizing the sensing waveform to match the target direction, where the predicted angle of target is needed. With the reasonable assumption that the target movement is slow, the target direction will not change significantly over the adjacent coherent time slots. Thus, the predicted angles are sufficient for the waveform optimization \cite{9652071}. This is a typical scenario in radar tracking, where the prior knowledge of the target direction is known for system design. Thus, the target angle $\vvarphi$ is assumed to be fixed in this study.

	\subsection{Channel Model}
	
	Consider the well-known 3D Saleh-Valenzuela channel model \cite{mmWaveChann}, the mmWave channel matrix between the BS and the RIS reflecting elements can be expressed as
	\begin{equation}
		\mathbf H = \sqrt{\frac{N_tM}{N_{cl}N_{ray}}}\sum_{i=1}^{N_{cl}}\sum_{l = 1}^{N_{ray}}\beta_{il} \bm\alpha_r(\eta_{il}^r,\phi_{il}^r)\bm\alpha_t^H(\eta_{il}^t,\phi_{il}^t), 
		\label{SVchan}
	\end{equation}
	where $N_{cl}$ and $N_{ray}$ denote the number of clusters and the number of rays in each cluster, and $\beta_l$ denotes the gain of the corresponding path. $\eta_{il}^r$, $\theta_{il}^r$, $\eta_{il}^t$, $\theta_{il}^t$ denote the azimuth/elevation angles of arrival/departure (AoA/AoD) of the $l$-th ray in the $i$-th cluster. Since we assume the antenna arrays at the BS and RIS are deployed on the x-z and y-z plane, respectively, the RIS and BS steering vectors are expressed as
	\begin{equation}
    \begin{split}
		\bm\alpha_r(\eta_l^r,\theta_l^r) = \frac{1}{\sqrt{M}}&\left[1, \cdots, e^{-j\frac{2\pi}{\lambda}(M_{y}-1)\sin\eta\sin\phi d_r} \right]^T \\
        \otimes &\left[1, \cdots, e^{-j\frac{2\pi}{\lambda}(M_{z}-1)\cos\phi d_r} \right]^T, 
    \end{split}
    \end{equation}
    \begin{equation}
    \begin{split}
		\bm\alpha_t(\eta_l^t,\theta_l^t) = \frac{1}{\sqrt{Nt}} &\left[1, \cdots, e^{-j\frac{2\pi}{\lambda}(N_{tx}-1)\cos\eta\sin\phi d_0} \right]^T  \\
        \otimes &\left[1, \cdots, e^{-j\frac{2\pi}{\lambda}(N_{tz}-1)\cos\phi d_0} \right]^T, 
    \end{split}
	\end{equation}
	where $d_r$ and $d_0$ are the element spacing of RIS elements and transmit antenna elements at BS. The channel from the RIS to user $k$ can be modeled as the similar manner, with only transmit array response due to single receive antenna at users.

	\section{Proposed Algorithm for CRB Minimization}
	
	In this section, we propose a joint hybrid beamforming and RIS coefficients design problem to minimize the CRB for estimating the direction of arrival (DoA). Then, a joint optimization scheme is proposed to solve the non-convex problem. 
	
	The optimization problem can be formulated as
	\begin{align}
		\min_{\mathbf F_{RF}, \mathbf F_{BB}, \atop \mathbf W, \mathbf\Theta} &\tr(\crb(\vvarphi))  \label{P1obj} \\
		\sto\quad &r_k \ge R_{th},  \label{Cons_rate}\tag{\ref{P1obj}a} \\
		& \mathbf W = \Frf\Fbb, \label{Cons_HBF}\tag{\ref{P1obj}b} \\
		& || \mathbf W ||_F^2 \le P_0, \label{Cons_pow}\tag{\ref{P1obj}c} \\
		& |\Frf(n_t,n_r)| = 1, \label{Cons_unit_Frf}\tag{\ref{P1obj}d} \\
		& |\phi_m| = 1, \label{Cons_uni_theta} \tag{\ref{P1obj}e}
	\end{align}
	where constraint (\ref{Cons_rate}) is to guarantee that the received data rate of each user is no less than an acceptable threshold value, (\ref{Cons_pow}) is for the transmit power budget, (\ref{Cons_unit_Frf}) and (\ref{Cons_uni_theta}) are unit modulus constraints for the phase shifts at BS and RIS, respectively. However, (\ref{P1obj}) is difficult to solve due to the non-convex objective function and constraints (\ref{Cons_rate}), (\ref{Cons_unit_Frf}) and (\ref{Cons_uni_theta}), as well as the strongly coupled variables. The proposed solution to this intractable problem will be given in the next section.
	
	To deal with the intractable objective function, we first introduce an auxiliary variable $\mathbf\Omega \succ 0$, where $\mathbf \Omega \in \mathbb C^{2\times 2}$ is a symmetric matrix, and then we have the following \textbf{Proposition 2.}
	
	\emph{Proposition 2.} \emph{Solving the CRB minimization problem in (\ref{P1obj}) is equivalent as solving the following problem,}
	\begin{align}
		\min_{\mathbf F_{RF}, \mathbf F_{BB}, \atop  \mathbf W,\mathbf\Theta, \mathbf\Omega} &\tr(\mathbf\Omega^{-1})  \label{P2obj} \\
		\sto\quad & \begin{bmatrix} 	
			\mathbf J_{\tilde{\bm\varphi}\tilde{\bm\varphi}} - \mathbf\Omega   & \mathbf J_{\tilde{\bm\varphi}\tilde{\bm\alpha}}  \\
			\mathbf J_{\tilde{\bm\varphi}\tilde{\bm\alpha}}^T   & \mathbf J_{\tilde{\bm\alpha}\tilde{\bm\alpha}}  \\
		\end{bmatrix} \succeq 0, \label{Cons_Schur} \tag{\ref{P2obj}a} \\
		& \mathbf\Omega \succ 0, \label{Cons_PD_Omega}\tag{\ref{P2obj}b}  \\
		& (\ref{Cons_rate}),(\ref{Cons_HBF}),(\ref{Cons_pow}),(\ref{Cons_unit_Frf}),(\ref{Cons_uni_theta}), \notag
	\end{align}
	
	\emph{The proof is given in \textbf{Appendix C}.} 
	
	Since variables $\mathbf\Theta$ and $\mathbf W$ are coupled in the FIM elements and thus difficult to handle, we introduce auxiliary matrix variables $\mathbf Z$ and $\mathbf U$ such that
	\begin{gather}
		\mathbf Z = \mathbf\Theta\mathbf H\mathbf W, \label{Cons_Z} \\
		\mathbf U = \mathbf Z \mathbf Z^H. \label{Cons_U}
	\end{gather}
	
	Hence, the entries of the FIM $\mathbf J$ can be rewritten as
    \begin{small}
	\begin{equation}
		\JppU = \frac{2|\alpha|^2T}{\sigma_r^2}\mathcal Re\left\{\begin{bmatrix} 
			\tr\left(\hat{\mathbf A}_\eta\mathbf U\hat{\mathbf A}_\eta^H\right), & \tr\left(\hat{\mathbf A}_\phi\mathbf U\hat{\mathbf A}_\eta^H\right) \\
			\tr\left(\hat{\mathbf A}_\eta\mathbf U\hat{\mathbf A}_\phi^H\right), & \tr\left(\hat{\mathbf A}_\phi\mathbf U\hat{\mathbf A}_\phi^H\right)
		\end{bmatrix} \right\},
	\end{equation}
 \end{small}
	\begin{equation}
		\JpaU= \frac{2T}{\sigma_r^2}\mathcal Re\left\{\begin{bmatrix} \alpha^*\tr\left( \mathbf{A}\mathbf U\hat{\mathbf A}_\eta^H \right) \\
			\alpha^*\tr\left( \mathbf{A}\mathbf U\hat{\mathbf A}_\phi^H \right) \end{bmatrix}\otimes [1,j] \right\},
	\end{equation}
	\begin{equation}
		\JaaU= \frac{2T}{\sigma_r^2}\mathcal Re\left\{ [1,j]^H[1,j] \tr\left( \mathbf{A}\mathbf U\mathbf A^H \right) \right\}.
	\end{equation}
	
	
	Moreover, the quadratic equality constraint (\ref{Cons_U}) is equivalent to the following two constraints by applying Schur complement:
	\begin{gather}
		\begin{bmatrix}
			\mathbf U & \mathbf Z\\
			\mathbf Z^H & \mathbf I
		\end{bmatrix} \succeq 0, \label{Cons_LMI} \\
		\tr\left(\mathbf U - \mathbf Z\mathbf Z^H\right) \le 0, \label{Cons_trace_Z}
	\end{gather}
	With (\ref{Cons_Z}), we have $\mathbf z_k = \mathbf\Theta \mathbf H \mathbf w_k, \, k = 1,\dots,N_{RF}$, where $\mathbf Z = [\mathbf z_1,\cdots,\mathbf z_{N_{RF}}]$. Combining the IUI and the sensing interference, the received SINR of user $k$ can be rewritten as
	\begin{equation}
		\gamma_k = \frac{\left\vert\mathbf g_k^H\mathbf z_k\right\vert^2}{\sum_{j=1,j\neq k}^{N_{RF}}\left\vert\mathbf g_k^H\mathbf z_j\right\vert^2 + \sigma_0^2}.
	\end{equation}

	Accordingly, problem (\ref{P2obj}) can be equivalently transformed as
	\begin{align}
		&\min_{\mathbf F_{RF}, \mathbf F_{BB},  \mathbf W, \atop\mathbf\Theta, \mathbf\Omega,\mathbf Z,\mathbf U} \tr(\mathbf\Omega^{-1})  \label{P3obj} \\
		\sto\quad & |\mathbf g_k^H\mathbf z_k|^2 \ge \left(2^\Rth-1\right)\left( \sum_{j\neq k} |\mathbf g_k^H\mathbf z_j|^2 + \sigma_0^2\right), \label{Cons_z_SINR} \tag{\ref{P3obj}a}\\
		& (\ref{Cons_HBF}),(\ref{Cons_pow}),(\ref{Cons_unit_Frf}),(\ref{Cons_uni_theta}),(\ref{Cons_Schur}),(\ref{Cons_PD_Omega}),(\ref{Cons_Z}), (\ref{Cons_LMI}), (\ref{Cons_trace_Z}). \notag
	\end{align}
	However, it is still challenging to solve (\ref{P3obj}) and find a global optimal due to its non-convexity caused by the constraints and coupled variables. Hence, the penalty method is applicable to address the coupled equality constraints by integrating them into the objective function as follows.
	\begin{align}
		\min_{\mathbf F_{RF}, \mathbf F_{BB},  \mathbf W,\atop\mathbf\Theta,  \mathbf\Omega, \mathbf Z,\mathbf U} & h(\mathbf F_{RF}, \mathbf F_{BB},  \mathbf W,\mathbf\Theta,  \mathbf\Omega, \mathbf Z, \mathbf U) = \tr(\mathbf\Omega^{-1})  \nonumber \\
        &+ \frac{\rho}{2}\Bigg( \left\Vert\mathbf W-\Frf\Fbb\right\Vert_F^2 + \left\Vert \mathbf Z-\mathbf\Theta\mathbf H\mathbf W\right\Vert_F^2 \Bigg)  \label{P4obj} \\
		\sto\quad	& (\ref{Cons_pow}),(\ref{Cons_unit_Frf}),(\ref{Cons_uni_theta}),(\ref{Cons_Schur}),(\ref{Cons_PD_Omega}), (\ref{Cons_LMI}), (\ref{Cons_trace_Z}), (\ref{Cons_z_SINR}),\notag
	\end{align}
	where $\rho$ is the penalty coefficient causing a high cost for the violation of the equality constraints. Hence, problem (\ref{P4obj}) is equivalent to (\ref{P3obj}) when $\rho \to \infty$ \cite{Bertsekas0Nonlinear}. The penalty factor $\rho$ is first set to a small value to find a good starting point, then it is updated in the outer layer until sufficiently large.
	
	Although the problem is still non-convex, the variables are separable. In the inner layer, when the penalty factor $\rho$ is fixed, we propose a BCD based algorithm by dividing the variables into two blocks, namely $\mathcal G_1 = \left\{\mathbf W, \Fbb,\mathbf\Omega,\mathbf Z,\mathbf U\right\}$ and $\mathcal G_2 = \left\{\Frf, \mathbf\Theta\right\}$, and optimize the variables alternatively based on the two blocks.

	\subsection{Penalty-CCCP-based Hybrid and Digital Beamforming Optimization}
	
	In this subsection, we optimize the hybrid beamforming and digital beamforming matrices $\mathbf W$ and $\Fbb$ as well as the auxiliary variables in $\mathcal G_1$ with fixed variables in $\mathcal G_2$. Thus, the problem can be reformulated as
	\begin{align}
		\min_{\mathcal G_1} \;&\tr(\mathbf\Omega^{-1}) + \frac{\rho}{2}\Bigg( \left\Vert\mathbf W-\Frf\Fbb\right\Vert_F^2 + \left\Vert \mathbf Z-\mathbf\Theta\mathbf H\mathbf W\right\Vert_F^2 \Bigg)  \label{P5obj} \\
		\sto\quad	& (\ref{Cons_pow}),(\ref{Cons_Schur}),(\ref{Cons_PD_Omega}),  (\ref{Cons_LMI}), (\ref{Cons_trace_Z}), (\ref{Cons_z_SINR}).\notag
	\end{align}
	The main difficulty in solving (\ref{P5obj}) arises from the non-convex feasible set due to the constraints (\ref{Cons_trace_Z}) and (\ref{Cons_z_SINR}). For (\ref{Cons_trace_Z}), the left-hand-side is a difference of convex (DC) function. Thus, the penalty method is extended to penalty-CCCP algorithm, where the concave part of (\ref{Cons_trace_Z}) is linearized by its first order Taylor expansion.
	
	The DC function constraint (\ref{Cons_trace_Z}) is equivalent to
	\begin{equation}
		\tr\left(\mathbf U\right) \le \tr\left(\mathbf{ZZ}^H\right), \label{temp1}
	\end{equation}
	The lower bound of the right hand side of (\ref{temp1}) is approximated by taking the first-order Taylor expansion,
	\begin{equation}
		\tr\left(\mathbf{ZZ}^H\right) \ge 2\mathcal Re\left\{\tr\left(\mathbf Z^{(t)}\mathbf Z^H\right)\right\} - \tr\left(\mathbf Z^{(t)}\left(\mathbf Z^{(t)}\right)^H\right),
	\end{equation}
	where $(t)$ means the updated value in last iteration. Thus, (\ref{temp1}) can be approximated as
	\begin{equation}
		2\mathcal Re\left\{\tr\left(\mathbf Z^{(t)}\mathbf Z^H\right)\right\} - \tr\left(\mathbf Z^{(t)}\left(\mathbf Z^{(t)}\right)^H\right) \ge \tr\left(\mathbf U\right). \label{Cons_DC_linear}
	\end{equation}
	
	Then, for non-convex constraint (\ref{Cons_z_SINR}), the phase of $\mathbf g_k^H\mathbf z_k$ can be rotated with an arbitrary phase shift. Thus, (\ref{Cons_z_SINR}) can be transformed into a SOC constraint as
	\begin{equation}
    \begin{split}
		\mathcal Re\{\mathbf g_k^H\mathbf z_k\} \ge \sqrt{2_{th}^R-1} &\left\Vert \left[ \mathbf g_k^H\mathbf z_1, \dots,\mathbf g_k^H\mathbf z_{k-1},\mathbf g_k^H\mathbf z_{k+1},\right.\right. \\
        &\left.\left.\dots, \mathbf g_k^H\mathbf z_{N_{RF}}, \sigma_0   \right] \right\Vert_2,  \label{Cons_SoC} 
    \end{split}
    \end{equation}
    \begin{equation}
		\mathcal Im\{\mathbf g_k^H\mathbf z_k\} = 0. \label{Cons_Im}
	\end{equation}
	To this end, problem (\ref{P5obj}) can be approximated in each iteration as 
	\begin{align}
		\min_{\mathcal G_1  } \;&\tr(\mathbf\Omega^{-1}) + \frac{\rho}{2}\Bigg( \left\Vert\mathbf W-\Frf\Fbb\right\Vert_F^2 + \left\Vert \mathbf Z-\mathbf\Theta\mathbf H\mathbf W\right\Vert_F^2 \Bigg)  \label{P6obj} \\
		\sto\quad	& (\ref{Cons_pow}),(\ref{Cons_Schur}),(\ref{Cons_PD_Omega}),  (\ref{Cons_LMI}), (\ref{Cons_DC_linear}), (\ref{Cons_SoC}),(\ref{Cons_Im}),\notag
	\end{align}
	which is a convex problem. Starting at a feasible initial point, this problem can be efficiently solved by the existing convex toolbox such as CVX \cite{2004Convex} and the corresponding variables can be updated in each iteration. The proposed CCCP-based algorithm for solving the sub-problem (\ref{P5obj}) is summarized as \textbf{Algorithm 1}, and the convergence analysis is in \textbf{Appendix D}.
	
	\begin{algorithm}[t]  
		\caption{Proposed CCCP-based Algorithm for Hybrid Beamforming Design.}
		\begin{algorithmic}[1]  
			\State \textbf{Initialize} variables $ \mathbf W^{(0)}, \mathbf Z^{(0)}$. Set iteration index $t=0$.
			\Repeat
			\State Update $\mathbf W^{(t+1)},\,\mathbf\Omega^{(t+1)},\,\mathbf Z^{(t+1)},\mathbf U^{(t+1)}$ by solving problem (\ref{P6obj}).
			\Until{the value of the objective function converges.}
		\end{algorithmic} 
	\end{algorithm}
	
	%
	
	\subsection{CCM-based Joint Optimization on Analog Beamforming and RIS Phase Shifts}
	
	Given updated variables in $\mathcal G_1$ and $\mathcal G_2$, the sub-problem optimizing $\Frf$ and $\mathbf\Theta$ can be decomposed as
	\begin{align}
		\min_{\mathbf F_{RF},\mathbf\Theta} &\frac{\rho}{2}\Bigg( \left\Vert\mathbf W-\Frf\Fbb\right\Vert_F^2 + \left\Vert \mathbf Z-\mathbf\Theta\mathbf H\mathbf W\right\Vert_F^2 \Bigg)  \label{P7obj} \\
		\sto\quad	& (\ref{Cons_unit_Frf}),(\ref{Cons_uni_theta}),\notag
	\end{align}
	where the objective function with respect to the two variables are separable. With unit modulus constraints, $\Frf$ and $\mathbf\Theta$ can form two independent Riemannian manifolds as
	\begin{gather}
		\mathcal M_1 = \{\mathbf f_{RF} \in \mathbb C^{N_{RF}N_s\times 1}| |f_{RF,1}| = \cdots = |f_{RF,N_{RF}N_s}| = 1\},\\
		\mathcal M_2 = \{\bm\vartheta \in \mathbb C^{M \times 1}| |\vartheta_1| = \cdots = |\vartheta_M| = 1\},
	\end{gather}
	where $\mathbf f_{RF} = \vect(\Frf) \in \mathbb C^{NtN_{RF}\times 1}$, and $\bm\vartheta = \ddot{\diag}\{\mathbf\Theta\} \in \mathbb C^{M\times 1}$, where $\ddot{\diag}\{\mathbf X\}$ is defined as a vector whose elements are the diagonal elements of matrix $\mathbf X$.
	After a series of transformations, problem (\ref{P7obj}) can be decomposed as two sub-problems, 
	\begin{gather}
		\min_{\mathcal M_1}  \;f_1(\mathbf f_{RF}) = \mathbf f_{RF}^H\mathbf\Xi_1\mathbf f_{RF} - 2\mathcal Re\left\{\bm\delta_1^H\mathbf f_{RF}\right\}  +\tau_1, \label{MOp1}  \\
		\min_{\mathcal M_2}  \;f_2(\bm\vartheta) = \bm\vartheta^H\mathbf\Xi_2\bm\vartheta - 2\mathcal Re\left\{\bm\delta_2^H\bm\vartheta\right\}  +\tau_2,\label{MOp2}
	\end{gather}
	where
	\begin{align} 
		& \mathbf\Xi_1 = (\Fbb^T \otimes \mathbf I_{N_t})^H(\Fbb^T\otimes \mathbf I_{N_t}), \\  &\mathbf\Xi_2 = \breve{\text{diag}}\left\{\mathbf{HWW}^H\mathbf H^H\right\}, \\
		&\bm\delta_1 = \left(\vect({\mathbf W})^H\mathbf(\Fbb^T\otimes \mathbf I_{N_t})\right)^H, \\
        &\bm\delta_2 =\ddot{\diag}\{\mathbf Z\mathbf W^H\mathbf H^H\}, \label{Def_Delta}\\
		&\tau_1 = \tr(\mathbf W\mathbf W^H), \; \tau_2 = \tr(\mathbf Z\mathbf Z^H),  \label{Def_tau} \end{align}
	where we define $\breve{\rm{\diag}}(\mathbf X)$ to denote a diagonal matrix whose diagonal elements are composed of the diagonal elements of a square matrix $\mathbf X$.
	
	\emph{Proof}: Because $\rho$ is a fixed constant in the inner group, we only focus on $f(\Frf,\mathbf\Theta) =  \left\Vert\mathbf W-\Frf\Fbb\right\Vert_F^2 + \left\Vert \mathbf Z-\mathbf\Theta\mathbf H\mathbf W\right\Vert_F^2$ in (\ref{P7obj}).
	
	The first term of $f(\Frf,\mathbf\Theta)$ can be transformed as
	\begin{equation}
		\begin{split}
			&\left\Vert\mathbf W-\Frf\Fbb\right\Vert_F^2 \\
			=\;&\tr\left( (\mathbf W- \Frf\Fbb)^H(\mathbf W- \Frf\Fbb) \right) \\
			=\; &\tr(\Fbb^H\Frf^H\Frf\Fbb)  \\
            &- 2\mathcal Re\{\tr(\mathbf W^H\Frf\Fbb)\} +\tr(\mathbf W^H\mathbf W)   \\
			=\;& \vect(\Frf)^H(\Fbb^T \otimes \mathbf I_{N_t})^H(\Fbb^T\otimes \mathbf I_{N_t})\vect(\Frf) \\
			&- 2\mathcal Re\{\vect({\mathbf W})^H\mathbf(\Fbb^T\otimes \mathbf I_{N_t})\vect(\Frf)\} +\tr(\mathbf W\mathbf W^H).
		\end{split} 
	\end{equation}
	The second term of $f(\Frf,\mathbf\Theta)$ can be transformed as
	\begin{equation}
		\begin{split}
			&\left\Vert \mathbf Z-\mathbf\Theta\mathbf H\mathbf W\right\Vert_F^2 \\
			=\; & \tr\left( (\mathbf Z-\mathbf\Theta\mathbf H\mathbf W)^H(\mathbf Z-\mathbf\Theta\mathbf H\mathbf W) \right) \\
			=\; & \tr(\mathbf\Theta\mathbf H\mathbf W\mathbf W^H\mathbf H^H\mathbf\Theta^H)  - 2\mathcal Re\{\tr(\mathbf Z^H\mathbf{\Theta HW})\} + \tr(\mathbf Z^H\mathbf Z) \\
			=\;&\bm\vartheta^H\breve{\text{diag}}\left\{\mathbf{HWW}^H\mathbf H^H\right\}\bm\vartheta - 2\mathcal Re\{\ddot{\diag}\{\mathbf Z\mathbf W^H\mathbf H^H\}^H \bm\vartheta\} \\
            &+ \tr(\mathbf Z\mathbf Z^H).
		\end{split}
	\end{equation}
	Because the objective function is separable with the two variables, the objective function $f(\Frf,\mathbf\Theta)$ can be rewritten as the form in (\ref{MOp1}) and (\ref{MOp2}), which can be directly solved by the CCM algorithm. The core of the CCM algorithm is to adopt the efficient gradient descent algorithm on the Riemannian manifold space. Since the two sub-problems are in the same form, the processes of the CCM optimization algorithm to solve them are similar, containing four main steps follows:
	\begin{enumerate}
		\item Derive the Euclidean gradient: To begin with, we compute the Euclidean gradient of $f(\bm\varpi)$ with respect to $\bm\varpi$ as
		\begin{equation}
			\nabla_{\bm\varpi}f_i(\bm\varpi) = 2\mathbf\Xi_i\bm\varpi - 2\bm\delta_i, \label{Euc_grad}
		\end{equation} 
		where $\bm\varpi \in \{\mathbf f_{RF}, \bm\vartheta\}$, $i \in \{1,2\}$.
		
		\item Derive the Riemannian gradient: The Riemannian gradient can be obtained by projecting the Euclidean gradient onto the tangent space $T_{\bm\varpi}\mathcal M$ of the manifold $\mathcal M$ at the current point $\bm\varpi$, where the tangent space is defined as
		\begin{equation}
        \begin{split}
			T_{\bm\varpi}\mathcal M_i =& \left\{\mathbf u_i\in\mathbb C^{D\times 1}: \mathcal Re\{\mathbf u_i\odot\bm\varpi_i^* \} = \mathbf 0 \right\}, \\
            &D \in \{N_tN_{RF}, M\}.
        \end{split}
		\end{equation}
		The Riemannian gradient of $f(\bm\varpi)$ can be expressed as
		\begin{equation}
			\grad_{\bm\varpi} f_i(\bm\varpi) = \nabla_{\bm\varpi}f_i(\bm\varpi) - \mathcal Re\{\left(\nabla_{\bm\varpi}f_i(\bm\varpi)\right)^*\odot \bm\varpi\}\odot\bm\varpi. \label{Riem_grad}
		\end{equation}
		
		\item Update the current point $\bar{\bm\varpi}$ over the tangent space: The updated point can be written as 
		\begin{equation}
			\bar{\bm\varpi} = \bvpi - \lambda \grad_{\bvpi} f_i(\bm\varpi), \label{Update_tangent}
		\end{equation}
		where $\lambda >0$ is a constant step size selected as the Polak-Ribiere parameter for fast convergence \cite{Polak}.
		
		\item Retraction: Note that the updated point on the tangent space may leave the manifold space $\mathcal M_i$. Thus, we need to retrieve the updated point into $\mathcal M_i$ as follows,
		\begin{equation}
			\bvpi^{t+1} = \bar{\bvpi} \odot \frac{1}{|\bar{\bvpi}|}. \label{Retrac}
		\end{equation} 
	\end{enumerate}
	
	Note that the problems (\ref{MOp1}) and (\ref{MOp2}) are unconstrained convex problems over the two manifolds. Thus, according to \cite{Polak}, the proposed MO algorithm is guaranteed to converge to the optimal point and is summarized in \textbf{Algorithm 2}.
	
	\begin{algorithm}[t]  
		\caption{Proposed MO Algorithm for Analog Beamforming and RIS Phase shifts Optimization.}
		\begin{algorithmic}[1]  
			\State \textbf{Initialize} the starting point $\bvpi^0$. Set the accuracy $\epsilon$. Set iteration index $t=0$.
			\Repeat
			\State Compute the Euclidean gradient $\nabla_{\bvpi}f_i(\bvpi)$ at the current point using (\ref{Euc_grad}).
			\State Compute the Riemannian gradient $\grad_{\bvpi} f_i(\bvpi)$ by orthogonal projection based on (\ref{Riem_grad}).
			\State Update the point $\bvpi^{t+1}$ on the tangent space based on (\ref{Update_tangent}).
			\State Retrieve the point onto the manifold space $\mathcal M_i$ as (\ref{Retrac}).
			\State Set the iteration index $t = t+1$.
			\Until{$|f_i(\bvpi^{t+1}) -  f_i(\bvpi^{t}) | < \epsilon$}
			\State \textbf{Output}: optimal $\Frf$, $\mathbf\Theta$.
		\end{algorithmic} 
	\end{algorithm}
	
    \subsection{Overall Convergence Analysis}
	
	In this part, we discuss about the convergence analysis of the proposed BCD-based algorithm. First, the overall algorithm solving (\ref{P1obj}) is summarized in \textbf{Algorithm 3}. 
	
	\begin{algorithm}[t]  
		\caption{Proposed BCD-based Algorithm for Joint Hybrid Beamforming, Radar Waveform and RIS Phase shifts Optimization Problem (\ref{P4obj})}.
		\begin{algorithmic}[1]  
			\State \textbf{Initialize} $\Frf^{(0)}$, $\Fbb^{(0)}$, $\mathbf W^{(0)}$, $\mathbf\Theta^{(0)}$, $\mathbf\Omega^{(0)}$, $\mathbf Z^{(0)}$, $\mathbf U^{(0)}$. Set iteration index $i=0$. Initialize penalty factor $\rho$.
			\Repeat:$\;$ outer loop
            \Repeat:$\;$ inner loop
			\State Update $\mathbf W^{(i+1)}$, $\Fbb^{(i+1)}$, $\mathbf\Omega^{(i+1)}$, $\mathbf Z^{(i+1)}$ and $\mathbf U^{(i+1)}$ based on \textbf{Algorithm 1}.
			\State Update $\Frf^{(i+1)}$ by \textbf{Algorithm 2}.
			\State Update $\mathbf\Theta^{(i+1)}$ by \textbf{Algorithm 2}.
			\State Set $i = i+1$.
			\Until{Convegence of the objective function (\ref{P4obj}.}
            \State Update $\rho$.
            \Until{$\rho$ is sufficiently large.}
			\State \textbf{Output}: optimal $\Frf$, $\Fbb$, $\mathbf W$, $\mathbf\Theta$.
		\end{algorithmic} 
	\end{algorithm}

	The prerequisite of discussing the convergence of \textbf{Algorithm 3} is that the algorithms for each sub-problems converges to a stationary point, which has been proved. This means each step of updating in \textbf{Algorithm 3} results in an local optimal solution to the corresponding sub-problem. Thus, the objective function $h(\mathbf F_{RF}, \mathbf F_{BB},  \mathbf W,\mathbf\Theta,  \mathbf\Omega, \mathbf Z, \mathbf U)$ monotonically decreases in each iteration update, shown as
	\begin{equation}
		\begin{split}
			&h\left(\mathbf F_{RF}^{(i+1)}, \mathbf F_{BB}^{(i+1)},  \mathbf W^{(i+1)},\mathbf\Theta^{(i+1)},  \mathbf\Omega^{(i+1)}, \mathbf Z^{(i+1)}, \mathbf U^{(i+1)}\right) \\ 
			&\overset{(a)}{\le} h\left(\mathbf F_{RF}^{(i)}, \mathbf F_{BB}^{(i+1)},  \mathbf W^{(i+1)},\mathbf\Theta^{(i)},  \mathbf\Omega^{(i+1)}, \mathbf Z^{(i+1)}, \mathbf U^{(i+1)}\right) \\
			&\overset{(b)}{\le} h\left(\mathbf F_{RF}^{(i)}, \mathbf F_{BB}^{(i)},  \mathbf W^{(i)},\mathbf\Theta^{(i)},  \mathbf\Omega^{(i)}, \mathbf Z^{(i)}, \mathbf U^{(i)}\right),
		\end{split}
	\end{equation}
	where $(a)$ is due to the update of $\mathbf F_{RF}^{(i+1)}$ and $\mathbf\Theta^{(i+1)}$, and $(b)$ is due to the update of $\mathbf W^{(i+1)}$,$F_{BB}^{(i+1)}$, $ \mathbf\Omega^{(i+1)}$, $\mathbf Z^{(i+1)}$ and $\mathbf U^{(i+1)}$. Thus, \textbf{Algorithm 3} converges to a local optimum for solving (\ref{P4obj}). Since solving (\ref{P1obj}) is equivalent to solving (\ref{P4obj}), the solution to the original problem minimizing the CRB for estimating the target direction (\ref{P1obj}) converges.
	
	\subsection{Computational Complexity Analysis}

    In this part, we analyze the computational complexity of the proposed \textbf{Algorithm 3}, which depends on \textbf{Algorithm 1} and \textbf{Algorithm 2}. In \textbf{Algorithm 1}, the total number of variables for solving problem (\ref{P6obj}) is $N_tN_s + N_tN_{RF} + N_{RF}N_s + MN_s + M^2 + 4$. Thus, the computational complexity of \textbf{Algorithm 1} is $\mathcal O(I_1\max(N_tN_{RF}, M^2 )^{3.5}\log_2(1/\epsilon))$, where $I_1$ is the number of iterations solving problem (\ref{P6obj}), and $\epsilon$ denotes the accuracy of the SCA algorithm. The computational complexity of \textbf{Algorithm 2} mainly depends on computing the Euclidean gradient of the two objective functions with respect to $\mathbf f_{RF}$ and $\bm\vartheta$ as (\ref{Euc_grad}), respectively. The complexity of computing the two gradients are $\mathcal O(N_t^2N_{RF}^2)$ and $\mathcal O(M^2)$. Hence, the  complexity of the two MO algorithm is $\mathcal O(I_2N_t^2N_{RF}^2\frac{1}{\epsilon^2} + I_3M^2\frac{1}{\epsilon^2})$, where $I_1$ and $I_2$ represent the corresponding numbers of iterations. Based on the above analysis, the total computational complexity of the proposed BCD-based \textbf{Algorithm 3} is $\mathcal O(ST(I_1\max(N_tN_{RF}, M^2 )^{3.5}\log_2(1/\epsilon) + I_2N_t^2N_{RF}^2\frac{1}{\epsilon^2} + I_3M^2\frac{1}{\epsilon^2}))$, where $S$ and $T$ denotes the numbers of iterations for updating the penalty factor $\rho$ and the BCD algorithm. 
	
	\section{Numerical Results}
		\begin{figure}[t]
		\centering
		\subfigure[CRB versus the number of SCA iterations in \textbf{Algorithm 1}.]{
			\begin{minipage}[t]{0.9\linewidth}
				\centering
				\includegraphics[width=1\linewidth]{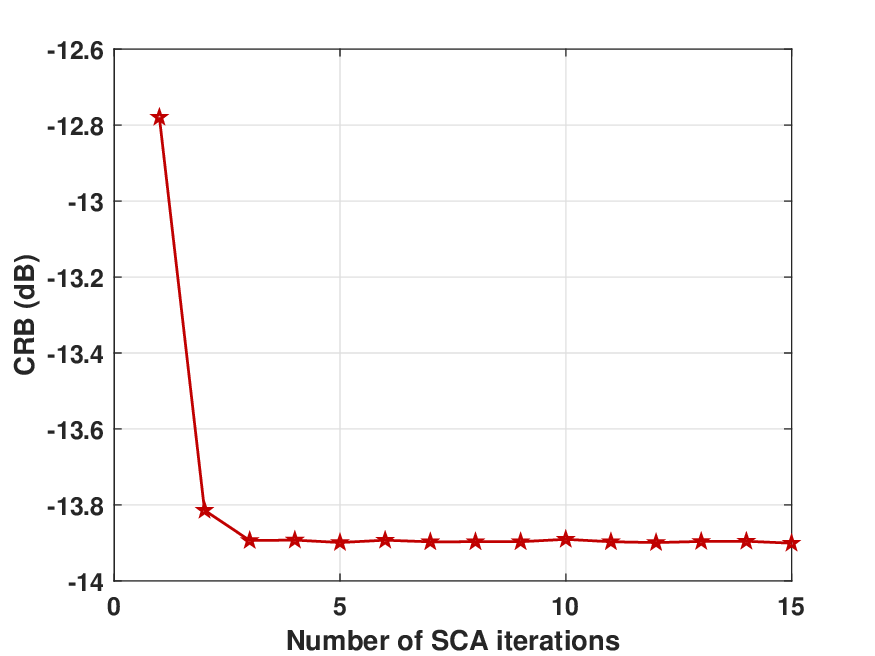}
				\label{Conv_alg1}
			\end{minipage}
		}
		\subfigure[CRB versus the number of BCD iterations in  \textbf{Algorithm 3}.]{
			\begin{minipage}[t]{0.9\linewidth}
				\centering
				\includegraphics[width=1\linewidth]{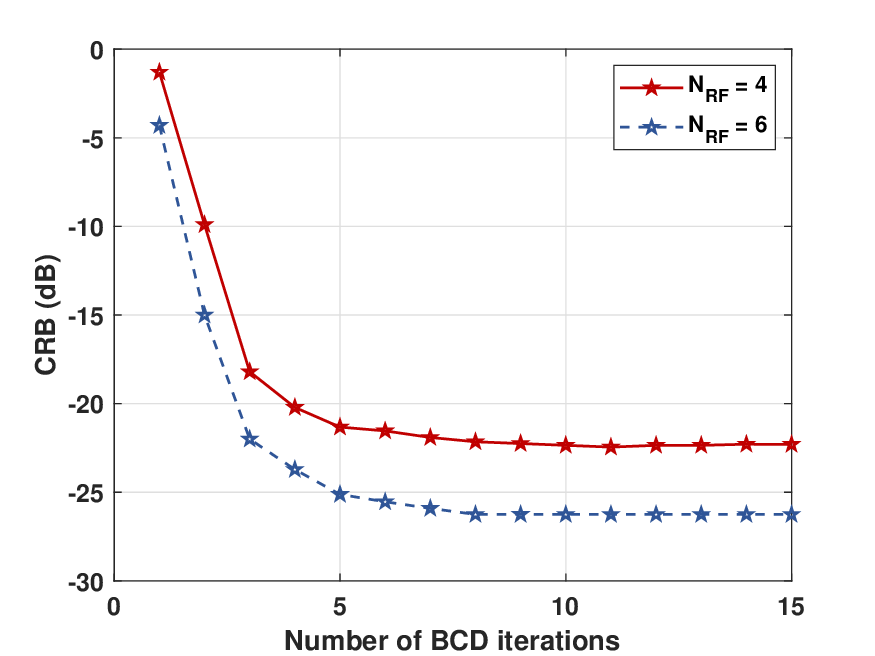}
				\label{Conv_alg3}
			\end{minipage}
		}
		\caption{Convergence performance analysis. Transmit power $P_0 = 30$ dBm, data rate threshold $R_{th} = 0.3$ bps/Hz, $M=50$ reflecting elements and $M_z=20$ sensing elements.}
		\label{Conv}
	\end{figure}
 
	In this section, the simulation results are provided to show the effectiveness of the proposed algorithm for joint sensing and communications design. The mmWave ISAC system works at 28 GHz. The BS is equipped with $N_t = 16\times 8$ UPA with $N_{RF} = 4$ or $N_{RF} = 6$ RF chains. The number of users is assumed to be $K = 3$. The noise powers are set to be $\sigma_0^2 = \sigma_r^2 = 90$ dBm. The distance between the BS and the RIS is set as $10$ m. \red{The distances between the RIS and the 3 users are set as 30m, and the angles of the users are randomly generated in the coverage area of the RIS.} The azimuth and elevation angles of the sensing target with respect to the RIS are set as $15^\circ$ and $75^\circ$, respectively. Assume that one coherent time block contains $T = 128$ symbols. \red{The SV channel model is considered as in (\ref{SVchan}), where the azimuth and elevation angles of the NLoS paths are randomly generated. The large scale fading factors $\beta_{il}$ satisfy Gaussian distribution such that $\beta_{il} \sim \mathcal CN(0, 10^{-0.1PL})$}, and the path loss can be computed as
	\begin{equation}
		PL = \gamma_a + 10\gamma_b\log_{10}(d) + \xi \text{ dB},
	\end{equation}
	where $d$ is the distance. According to \cite{6834753}, $\gamma_a = 61.4$, $\gamma_b = 2$, $\xi = 5.8$ dB for the line-of-sight (LoS) path, and $\gamma_a = 72.0$, $\gamma_b = 2.92$, $\xi = 8.7$ dB for non-line-of-sight (NLoS) paths.
	
	To show the effectiveness of the proposed algorithm, we compare the performance with other baselines, i.e. i) full digital beamforming with $N_{RF} = N_t$ RF chains; ii) alternating hybrif beamforming (AHBF) as in \cite{9729809}, where the beamforming matrix is first optimized as full digital case, then $\Fbb$ and $\Frf$ are alternatively optimized to approach the optimal digital beamforming matrix; iii) random phase configuration. 
	
	\subsection{Convergence Verification}
	
	To show the convergence performance of the proposed algorithm, Fig. \ref{Conv_alg1} and Fig. \ref{Conv_alg3} illustrate the achieved CRB with the number of iterations in SCA based \textbf{Algorithm 1} and BCD based \textbf{Algorithm 3}, respectively. From Fig. \ref{Conv_alg1}, the achieved CRB with \textbf{Algorithm 1} decreases and converges to a stable value in very few iterations. Given the output of \textbf{Algorithm 1}, the MO algorithm for analog beamforming and RIS phase shifts design reaches better estimation accuracy. As for the overall algorithm, We can observe from Fig. \ref{Conv_alg3} that the CRB for target direction estimation decreases gradually as the number of BCD iterations increases, and then converges to an optimum value within several iterations.

	\subsection{CRB versus Transmit Power Budget}
	
	\begin{figure}[t]
		\centering
		\includegraphics[width=0.9\linewidth]{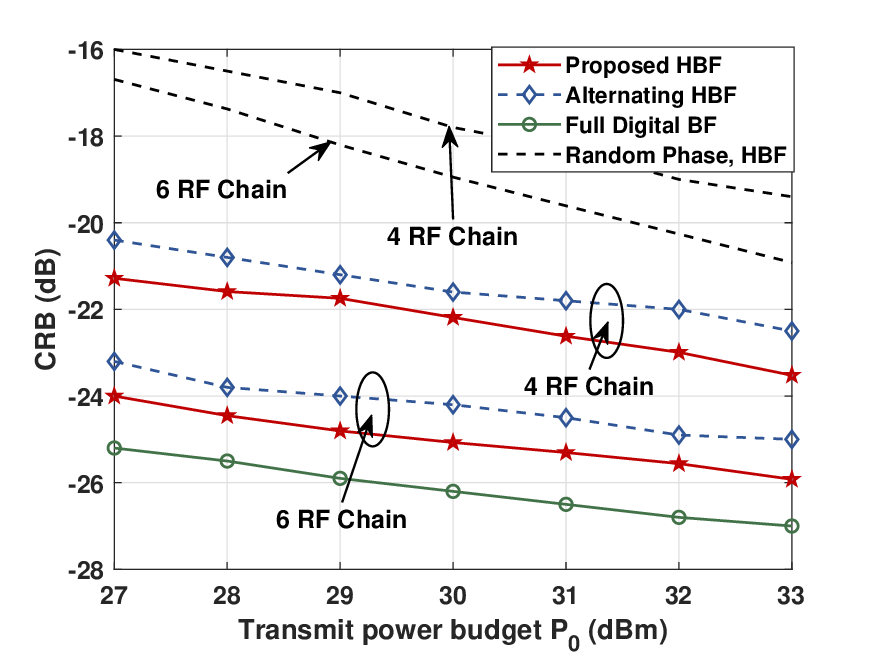}
		\caption{CRB versus transmit power budget $P_0$. Data rate threshold $R_{th} = 0.3$ bps/Hz, $M=50$ reflecting elements and $M_z = 20$ sensing elements.}
		\label{CRBvsP0}
	\end{figure}

	In Fig. \ref{CRBvsP0}, we analyze how CRB varies with transmit power budget $P_0$. When the transmit power at the BS rises from $27$ dBm to $33$ dBm, decrease on CRB can be observed accordingly. This is because larger transmit power at the BS results in larger illumination power at the sensing target, as well as stronger received signal at the users. \red{Nevertheless, energy efficiency is usually a key indicator in practical scenarios. How to balance the power consumption and both communication and sensing performance is also an indispensable research topic}. Compared with random phase shifts at RIS, the proposed algorithm and other baselines can reach remarkable CRB reduction, verifying the effectiveness of RIS phase shifts design. In addition, the proposed algorithm outperforms the alternating HBF method considerably, because alternating HBF first optimize the HBF matrix $\mathbf W$, and then decompose it as analog and digital beamforming matrices, which will introduce performance loss. Moreover, HBF with 6 RF chains significantly improves the sensing performance compared with 4 RF chains, approaching full digital beamforming (with 128 RF chains). Thus, trade-off between system performance and hardware cost must be taken into consideration in practical engineering design.
	
	\subsection{CRB versus the number of RIS Reflecting Elements}
	
	\begin{figure}[t]
		\centering
		\includegraphics[width=0.9\linewidth]{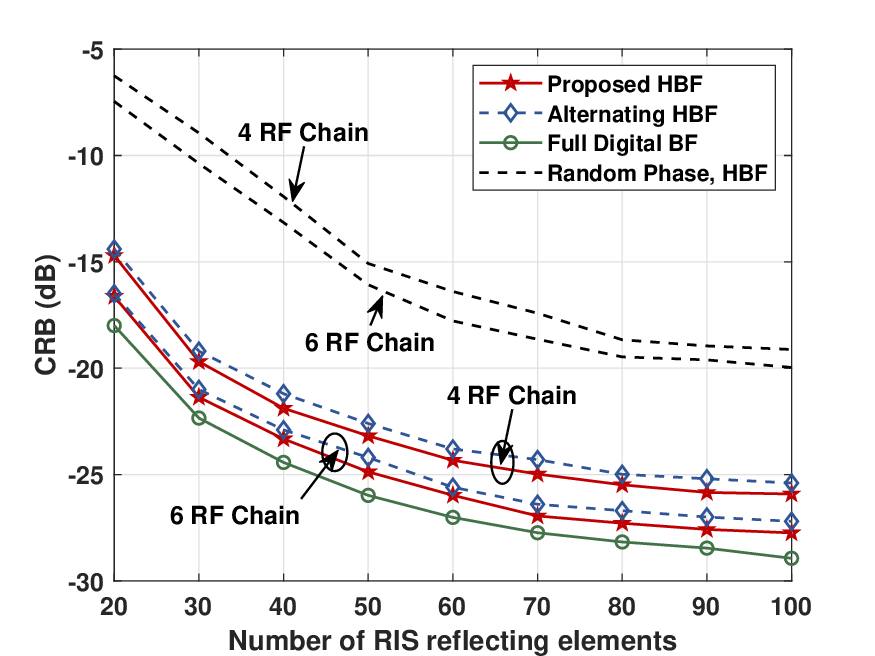}
		\caption{CRB versus the number of reflecting elements. Transmit power $P_0 = 30$ dBm. Data rate threshold $R_{th} = 0.3$ bps/Hz, and $M_z = 20$ sensing elements.}
		\label{CRBvsRefl}
	\end{figure}
	
	Fig. \ref{CRBvsRefl} shows the CRB performance versus the number of reflecting elements $M$ configured at the RIS. \red{It is true that larger number of elements on RIS causes higher hardware cost, computational complexity, as well as more power consumption to maintain the reflecting elements. Despite this, we obtain significant decrease of CRB with $M$ growing from 20 to 100, since larger number of reflecting elements provides more degrees of freedom to reconfigure the electromagnetic environment, and the energy utilization can be improved with stronger beams towards users and target}. Increasing the number of RF chains at the BS also contributes to the sensing accuracy by providing larger beamforming gain towards the sensing target, despite higher cost caused by the hardware implementations at the RF chains. Fortunately, the proposed HBF algorithm achieves an impressive sensing performance, achieving nearly 96\% of the accuracy achieved by full digital beamforming, as measured by the CRB. This offers a viable alternative to the challenging hardware design requirements of mmWave systems.

	\subsection{CRB versus the number of RIS Sensing Elements}
	
	\begin{figure}[t]
		\centering
		\includegraphics[width=0.9\linewidth]{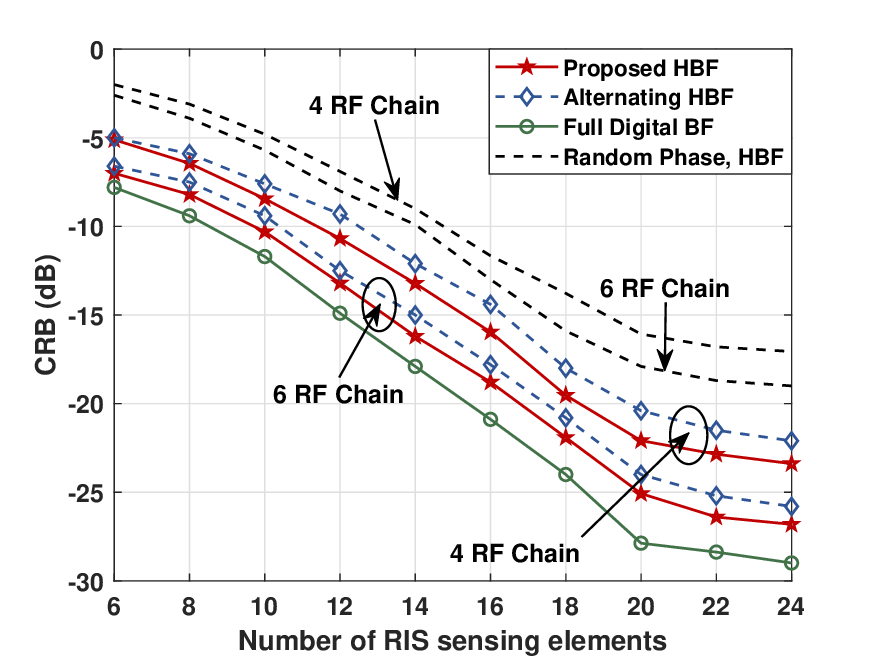}
		\caption{CRB versus the number of sensing elements.}
		\label{CRBvsSensing}
	\end{figure}

	The impact of the number of RIS reflecting elements on estimation accuracy is illustrated in Fig. \ref{CRBvsSensing}. It can be observed that the CRB drops with more sensing elements configured on RIS, and reaches to a stationary point. More sensing elements result in larger dimension on echo signal, improving the estimation accuracy. Also, the transmit array with 6 RF chains can achieve better performance than 4 RF chains, narrowing the performance gap between hybrid beamforming and digital beamforming. 
	
	\subsection{Trade-off Between Sensing and Communication}
	
	\begin{figure}[t]
		\centering
		\includegraphics[width=0.9\linewidth]{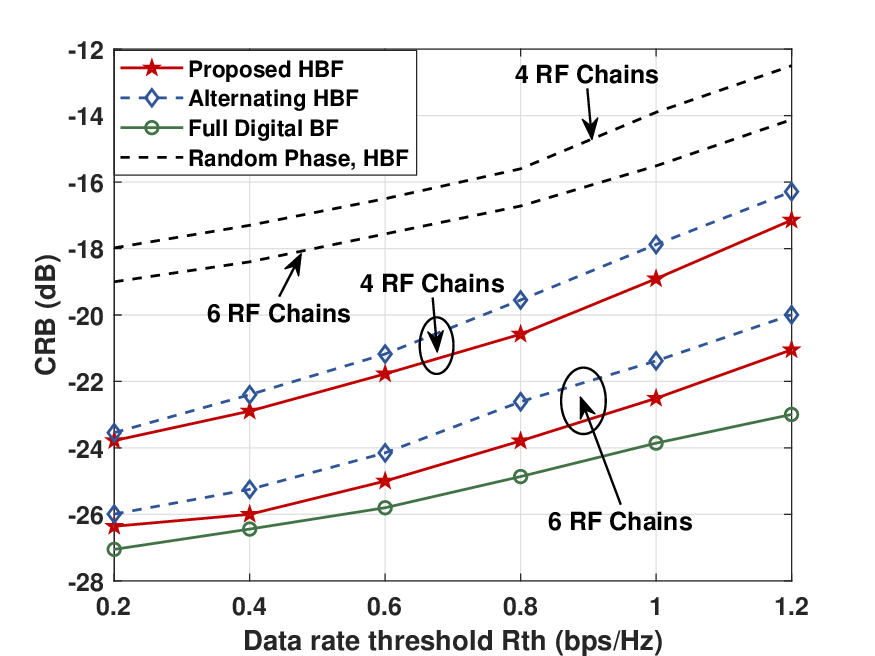}
		\caption{CRB versus the data rate threshold. Transmit power $P_0 = 30$ dBm, $M=50$ reflecting elements and $M_z=20$ sensing elements.}
		\label{CRBvsRth}
	\end{figure}
	The trade-off between sensing and communications performance is analysis and shown as in Fig. \ref{CRBvsRth}. To cater for higher communications demands, we increase the data rate threshold from 0.2 bps/Hz to 1.2 bps/Hz, and then the sensing performance deteriorates in terms of the CRB. This is intuitive because higher data rate threshold requires more beam gain allocated to the communication users, and sensing performance loss is unavoidable accordingly, given the same system configurations. \red{Thus, in practical engineering, how to balance the performance of communication and sensing in accordance with the real-time demands is of great importance.} Compared with the baselines, significant sensing performance improvement by the proposed algorithm can be also witnessed as in the former results. 
	
	\subsection{MSE Performance Verification}
   
\begin{figure}[t]
		\centering
        \includegraphics[width=0.9\linewidth]{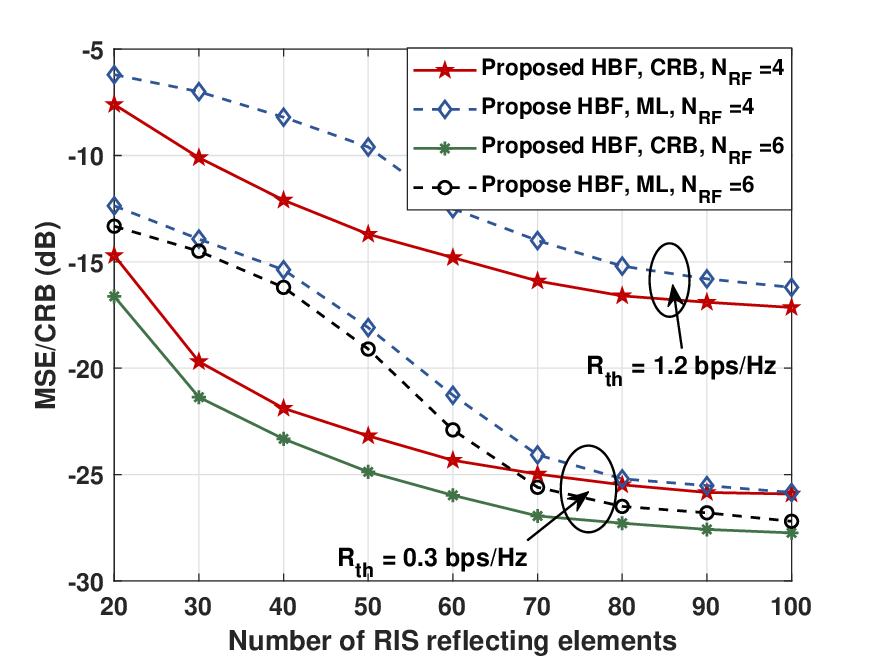}
		\caption{MSE/CRB versus the number of RIS reflecting elements. Transmit power $P_0 = 30$ dBm, and $M_s=20$ sensing elements.}
		\label{MSEvsM}
	\end{figure}

\begin{figure}[t]
		\centering
		\includegraphics[width=0.9\linewidth]{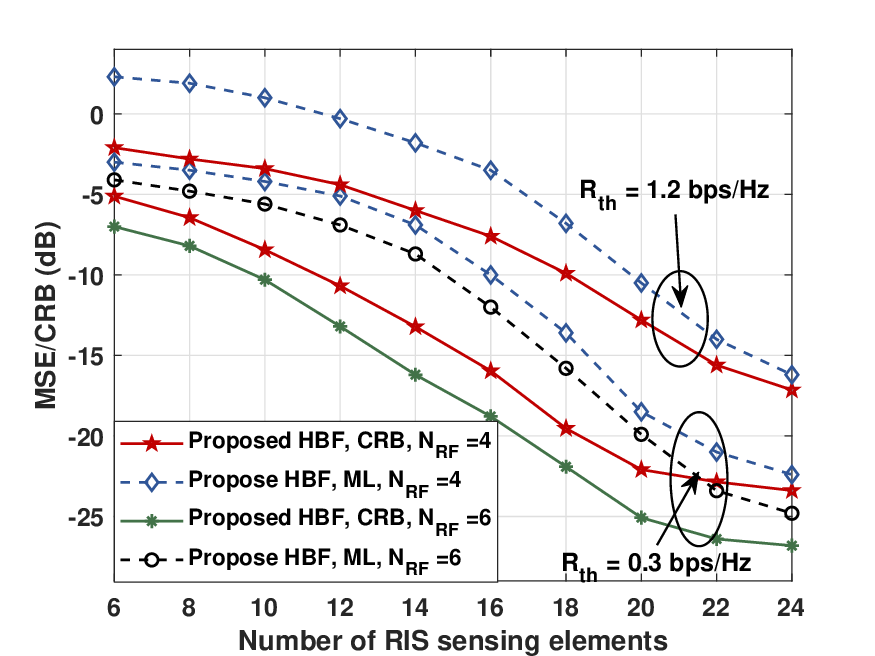}
		\caption{MSE/CRB versus the number of RIS sensing elements. Transmit power $P_0 = 30$ dBm, and $M=50$ reflecting elements.}
		\label{MSEvsMs}
	\end{figure}

    Next, to show that our CRB minimization algorithm is effective for improving estimation accuracy, the maximum likelihood (ML) estimation is employed with the optimal beamforming design, where the detail is in \textbf{Appendix D}. With ML estimation, the MSE of the azimuth and elevation angles of the sensing target is evaluated in Fig. \ref{MSEvsM} and $\ref{MSEvsMs}$. \red{The results verify that MSE is always lower bounded by CRB, and the gap is gradually narrowed with more reflecting/sensing elements.} Specifically, in Fig. \ref{MSEvsM}, the MSE / CRB versus the number of reflecting elements $M$ is illustrated. It can be observed that both the CRB and the MSE monotonically decrease with enhanced RIS beamforming gain. Moreover, when $M$ becomes larger, the MSE for ML estimation approaches the CRB, which is consistent with the results presented in \cite{1993Fundamentals, 2014Fundamentals, RISsensing}. With the same data rate threshold ($R_{th} = 0.3$ bps/Hz), hybrid beamforming with 6 RF chains also outperforms 4 RF chains in terms of estimation MSE. We further set $R_{th} = 1.2$ bps/Hz to meet more stringent communications demands. The results reveal the trade-off between the sensing and communication functions, due to the sensing accuracy performance loss with $R_{th} = 1.2$ bps/Hz. Fig. \ref{MSEvsMs} demonstrates the MSE / CRB versus the number of sensing elements $M_s$. Consistent the former results, the angle estimation accuracy is improved with more sensors at the RIS. \red{From the results, it is effective to improve the estimation accuracy by minimizing the corresponding CRB.}

 
	\section{Conclusion}
	
	This paper investigated a mmWave ISAC system aided by a semi-self sensing RIS. The CRB for DoAs estimation was minimized through hybrid beamforming and RIS reflecting coefficients design, where the communication data rate was guaranteed in the constraints. An efficient algorithm was proposed to tackle the non-convexity and the couples variables. In the simulation results, the convergence performance for the SCA-based algorithm and the overall algorithm was verified, which was consistent with the theoretical analysis. The significant performance improvement compared with the random phase shifts case revealed the effectiveness of RIS beamforming. Moreover, the propose HBF algorithm outperformed the alternating HBF method in CRB minimization. The trade-off between sensing and communication was analyze, where the estimation accuracy monotonically deteriorated with increasing data rate threshold. The results also showed that more RF chains can provide better performance by beamforming, though the hardware cost must be taken into consideration in real practice. Finally, the comparison between theoretical CRB and practical MSE was presented, which further confirmed the effectiveness of the CRB minimization approach.

    \red{Furthermore, based on this work, more comprehensive and practical scenarios can be considered, such as near-field and extended target assumption, clutters existing environment and multi-RIS to better enhance both sensing and communcation performance. Also, user channel and target location estimation error should also be considered in real practice.}
	
	
	\appendices 
	\section{Derivation of Fisher Information Matrix for the point target}
	Recalling (\ref{FIM_elem}), we first derive the partial derivatives of $\mathbf p$ with respect to $\vvarphi$ and $\valpha$ as follows,
	\begin{gather}
		\frac{\partial\mathbf p}{\partial\vvarphi} = [\alpha\vect(\hat{\mathbf A}_\eta\mathbf{\Theta HX}),\,\alpha\vect(\hat{\mathbf A}_\phi\mathbf{\Theta HX})], \\
		\frac{\partial\mathbf p}{\partial\valpha} = [1,j]\otimes\vect(\mathbf{A\Theta HX}),
	\end{gather}
	where $\mathbf A = \mathbf a(\eta,\phi)\mathbf b^T(\eta,\phi)$. To derive the partial derivatives of $\mathbf A$ with respect to $\eta$ and $\phi$ (namely,  $\hat{\mathbf A}_\eta$ and $\hat{\mathbf A}_\phi$), we rewrite the steering vectors as follows,
	\begin{equation} 
		\mathbf a(\eta,\phi) =\frac{1}{\sqrt{M_s}} e^{-j\bm\nu_s},\; \mathbf b(\eta,\phi) = \frac{1}{\sqrt{M}}e^{-j\bm\nu}
	\end{equation}
	where $\bm\nu_s = \frac{2\pi}{\lambda}(\bm\mu_{sY}\sin\eta\sin\phi d_y + \bm\mu_{sZ}\cos\phi d_z)$, $\bm\nu = \frac{2\pi}{\lambda}(\bm\mu_Y\sin\eta\sin\phi d_y + \bm\mu_Z\cos\phi d_z)$. $\bm\mu_{sY}$ and $\bm\mu_{sZ}$ denote the element indices of sensing elements at $Y$ and $Z$ axes, respectively, and $\bm\mu_{Y}$ and $\bm\mu_{Z}$ denote those of reflecting elements. To this end, the partial derivatives of $\mathbf A$ can be expressed as
	\begin{equation}
		\begin{split}
			\hat{\mathbf A}_\eta &= \frac{\partial\mathbf A}{\partial \eta} = \frac{\partial\mathbf a(\eta,\phi)}{\partial\eta}\mathbf b^T(\eta,\phi) + \mathbf a(\eta,\phi)\frac{\partial \mathbf b^T(\eta,\phi)}{\partial\eta}  \\
			&= -j\frac{2\pi}{\lambda\sqrt{M_sM}}\cos\eta\sin\phi d_y \\
            &\times \left( \diag\{\bm\mu_{sY}\}\mathbf{ab}^T + \mathbf{ab}^T\diag\{\bm\mu_Y\}  \right),
		\end{split}
	\end{equation}
	\begin{equation}
		\begin{split}
			\hat{\mathbf A}_\phi &= \frac{\partial\mathbf A}{\partial \phi} = \frac{\partial\mathbf a(\eta,\phi)}{\partial\phi}\mathbf b^T(\eta,\phi) + \mathbf a(\eta,\phi)\frac{\partial \mathbf b^T(\eta,\phi)}{\partial\phi} \\
			&= -j\frac{2\pi}{\lambda\sqrt{M_sM}}\sin\eta\cos\phi d_y \\
            &\times \left( \diag\{\bm\mu_{sY}\}\mathbf{ab}^T + \mathbf{ab}^T\diag\{\bm\mu_Y\}  \right)  \\
			&\quad+ j\frac{2\pi}{\lambda\sqrt{M_sM}}\sin\phi d_z \\
            & \times \left( \diag\{\bm\mu_{sZ}\}\mathbf{ab}^T + \mathbf{ab}^T\diag\{\bm\mu_Z\}  \right).
		\end{split}
	\end{equation}
	With $\hat{\mathbf A}_\eta$ and $\hat{\mathbf A}_\phi$, the first block $\mathbf J_{\vvarphi\vvarphi}$ can be derived as (\ref{Jpp}) at the bottom of this page. Similarly, we derive the matrix blocks $\mathbf J_{\vvarphi\valpha}$ and $\mathbf J_{\alpha\valpha}$ as (\ref{Jpa}) and (\ref{Jaa}).
	
	Thus, each entry of (\ref{FIM}) can be computed by (\ref{Jpp}), (\ref{Jpa}) and (\ref{Jaa}), respectively.

 	\begin{figure*}[b]
	\hrule
    \begin{align}
			&\mathbf J_{\vvarphi\vvarphi} = \frac{2}{\sigma_r^2}\mathcal Re\left\{ \frac{\partial \mathbf p^H}{\partial\vvarphi}\frac{\partial \mathbf p}{\partial\vvarphi}\right \}=  \frac{2}{\sigma_r^2}\mathcal Re\left\{\begin{bmatrix}
				\alpha^* \vect\left( \hat{\mathbf A}_\eta\mathbf{\Theta HX} \right) ^H \\
				\alpha^* \vect\left( \hat{\mathbf A}_\phi\mathbf{\Theta HX} \right) ^H
			\end{bmatrix} [\alpha\vect(\hat{\mathbf A}_\eta\mathbf{\Theta HX}),\,\alpha\vect(\hat{\mathbf A}_\phi\mathbf{\Theta HX})]  \right\}  \nonumber \notag\\
			&\quad\;\;\;= \frac{2|\alpha|^2}{\sigma_r^2}\mathcal Re\left\{\begin{bmatrix} 
				\tr\left(\hat{\mathbf A}_\eta\mathbf{\Theta HX} \mathbf X^H\mathbf H^H\mathbf\Theta^H\hat{\mathbf A}_\eta^H\right), & \tr\left(\hat{\mathbf A}_\phi\mathbf{\Theta HX} \mathbf X^H\mathbf H^H\mathbf\Theta^H\hat{\mathbf A}_\eta^H\right) \\
				\tr\left(\hat{\mathbf A}_\eta\mathbf{\Theta HX} \mathbf X^H\mathbf H^H\mathbf\Theta^H\hat{\mathbf A}_\phi^H\right), & \tr\left(\hat{\mathbf A}_\phi\mathbf{\Theta HX} \mathbf X^H\mathbf H^H\mathbf\Theta^H\hat{\mathbf A}_\phi^H\right)
			\end{bmatrix} \right\}\nonumber \notag \\
			&\quad\;\;\;= \frac{2|\alpha|^2T}{\sigma_r^2}\mathcal Re\left\{\begin{bmatrix} 
				\tr\left(\hat{\mathbf A}_\eta\mathbf{\Theta H} \Rx\mathbf H^H\mathbf\Theta^H\hat{\mathbf A}_\eta^H\right), & \tr\left(\hat{\mathbf A}_\phi\mathbf{\Theta H} \Rx\mathbf H^H\mathbf\Theta^H\hat{\mathbf A}_\eta^H\right) \\
				\tr\left(\hat{\mathbf A}_\eta\mathbf{\Theta H} \Rx\mathbf H^H\mathbf\Theta^H\hat{\mathbf A}_\phi^H\right), & \tr\left(\hat{\mathbf A}_\phi\mathbf{\Theta H} \Rx\mathbf H^H\mathbf\Theta^H\hat{\mathbf A}_\phi^H\right)
			\end{bmatrix} \right\}. \label{Jpp}\\
			&\mathbf J_{\vvarphi\valpha} = \frac{2}{\sigma_r^2}\mathcal Re\left\{\begin{bmatrix}
				\alpha^* \vect\left( \hat{\mathbf A}_\eta\mathbf{\Theta HX} \right) ^H \\
				\alpha^* \vect\left( \hat{\mathbf A}_\phi\mathbf{\Theta HX} \right) ^H
			\end{bmatrix}\left([ 1, j]\otimes \vect\left( \mathbf{A\Theta HX} \right)\right)  \right\}\nonumber \notag \\
			&\quad\;\;\;= \frac{2}{\sigma_r^2}\mathcal Re\left\{\begin{bmatrix} \alpha^*\tr\left( \mathbf{A\Theta HXX}^H\mathbf H^H\mathbf \Theta^H\hat{\mathbf A}_\eta^H \right) \\
				\alpha^*\tr\left( \mathbf{A\Theta HXX}^H\mathbf H^H\mathbf \Theta^H\hat{\mathbf A}_\phi^H \right) \end{bmatrix}\otimes [1,j] \right\} = \frac{2T}{\sigma_r^2}\mathcal Re\left\{\begin{bmatrix} \alpha^*\tr\left( \mathbf{A\Theta H}\Rx\mathbf H^H\mathbf \Theta^H\hat{\mathbf A}_\eta^H \right) \\
				\alpha^*\tr\left( \mathbf{A\Theta H}\Rx\mathbf H^H\mathbf \Theta^H\hat{\mathbf A}_\phi^H \right) \end{bmatrix}\otimes [1,j] \right\}.
        \label{Jpa} \\
        &\mathbf J_{\valpha\valpha}  = \frac{2}{\sigma_r^2}\mathcal Re\left\{ \left( [1,j]\otimes\vect(\mathbf{A\Theta HX}) \right)^H \left(\right[1,j]\otimes\vect(\mathbf{A\Theta HX}))\right\} \nonumber \notag\\
			&\quad\;\;\; =\frac{2}{\sigma_r^2}\mathcal Re\left\{ [1,j]^H[1,j]\tr\left( \mathbf{A\Theta HXX}^H\mathbf H^H\mathbf\Theta^H\mathbf A^H \right) \right\}  = \frac{2T}{\sigma_r^2}\mathcal Re\left\{ [1,j]^H[1,j]\tr\left( \mathbf{A\Theta H}\Rx\mathbf H^H\mathbf\Theta^H\mathbf A^H \right) \right\}.\label{Jaa}
	\end{align}
    \end{figure*}
	

	\section{Proof of Proposition 2}
	
	To prove \textbf{Proposition 2}, first we have the FIM $\mathbf J$ is always a symmetric and positive semidefinite matrix, we then assume that $\mathbf J$ is nonsingular and thus positive definite \footnote{Note that if the FIM is a singular matrix, the unbiased estimator will not exist, and thus the CRB will be unbounded \cite{9652071}.}. For notation simplicity, we let $\mathbf S = \mathbf J_{\vvarphi\vvarphi} - \mathbf J_{\vvarphi\valpha}\mathbf J_{\valpha\valpha}^{-1}\mathbf J_{\vvarphi\valpha}^T \succ 0$.
	
	Because the matrix function $f(\mathbf X) = \tr(\mathbf X^{-1})$ is convex on $S_{++}^n$ \cite{2004Convex}, we have the first order condition $f(\mathbf S) \ge f(\mathbf\Omega) + \nabla f(\mathbf\Omega)(\mathbf S-\mathbf\Omega)$ for $\mathbf S\succ 0$, $\mathbf\Omega \succ 0$ and $\mathbf S-\mathbf\Omega \succeq 0$. Let $\mathbf U = \mathbf\Omega -\mathbf S$, the first order condition can be rewritten as
	\begin{equation}
		\begin{split}
			\tr(\mathbf \Omega^{-1}) &\ge \tr(\mathbf S^{-1}) + \lim\limits_{h\to 0}\frac{f(\mathbf S+\mathbf U) - f(\mathbf S)}{h} \\
			&= \tr(\mathbf S^{-1}) + \lim\limits_{h\to 0}\frac{\tr((\mathbf S+\mathbf U)^{-1}) - \tr(\mathbf S^{-1})}{h} \\
			& = \tr(\mathbf S^{-1}) + \tr\left(\lim\limits_{h\to 0}\frac{(\mathbf S+\mathbf U)^{-1} - \mathbf S^{-1}}{h} \right) \\
			& = \tr(\mathbf S^{-1}) + \tr\left(\lim\limits_{h\to 0} (\mathbf S+h\mathbf U)^{-1} \frac{\mathbf S - (\mathbf S + h\mathbf U)}{h}\mathbf S^{-1} \right) \\	
			& = \tr(\mathbf S^{-1})  - \tr(\mathbf S^{-1}\mathbf U\mathbf S^{-1}) \\
			& = \tr(\mathbf S^{-1})  + \tr(\mathbf S^{-2}\mathbf (\mathbf S - \mathbf \Omega )).
		\end{split}
	\end{equation}
	Since $\mathbf S^{-2}$ and $\mathbf S-\mathbf \Omega$ are both positive semidefinite matrices, then $\tr(\mathbf S^{-2}\mathbf (\mathbf S - \mathbf \Omega )) \ge 0$. Thus, we have $\tr(\mathbf\Omega^{-1}) \ge \tr(\mathbf S^{-1})$. Minimizing $\crb(\vvarphi) = \tr(\mathbf S^{-1})$ is equivalent to minimizing its upper bound $\tr(\mathbf\Omega^{-1})$ under the constraint on 
	\begin{equation}
		\mathbf S = \mathbf J_{\vvarphi\vvarphi} - \mathbf J_{\vvarphi\valpha}\mathbf J_{\valpha\valpha}^{-1}\mathbf J_{\vvarphi\valpha}^T \succeq \mathbf\Omega. \label{diff_PSD}
	\end{equation}
	According to the Schur complement, (\ref{diff_PSD}) is equivalent to
	\begin{equation}
		\begin{bmatrix} 	
			\mathbf J_{\tilde{\bm\varphi}\tilde{\bm\varphi}} - \mathbf\Omega   & \mathbf J_{\tilde{\bm\varphi}\tilde{\bm\alpha}}  \\
			\mathbf J_{\tilde{\bm\varphi}\tilde{\bm\alpha}}^T   & \mathbf J_{\tilde{\bm\alpha}\tilde{\bm\alpha}}  \\
		\end{bmatrix} \succeq 0.
	\end{equation}
	
	Thus, \textbf{Proposition 1} is proved.

	\section{Convergence Analysis of Algorithm 1}
	
	Considering constraints (\ref{temp1}) and (\ref{Cons_z_SINR}), we can define two functions as follows,
	\begin{gather}
		\mathcal Q_1(\mathbf Z) = \tr(\mathbf D_1) - \tr(\mathbf Z\mathbf Z^H), \\
		\mathcal Q_2(\mathbf z_k) = \left(2^\Rth-1\right)\left( \sum_{j\neq k} |\mathbf g_k^H\mathbf z_j|^2 + \sigma_0^2\right) - |\mathbf g_k^H\mathbf z_k|^2.
	\end{gather}
	Note that we always have $\mathcal Q_1(\mathbf Z) \le 0$ and $\mathcal Q_2(\mathbf z_k) \le 0$. In each iteration, define another two functions that replace $\tilde{\mathcal F}_1(\mathbf Z)$ and $\tilde{\mathcal F}_2(\mathbf z_k) \le 0$ as follows,
	\begin{align}
		\tilde{\mathcal Q}_1(\mathbf Z|\mathbf Z^{(t)}) &= 	\tr\left(\mathbf D_1\right) - 2\mathcal Re\left\{\tr\left(\mathbf Z^{(t)}\mathbf Z^H\right)\right\} \nonumber \notag \\
        &+ \tr\left(\mathbf Z^{(t)}\left(\mathbf Z^{(t)}\right)^H\right),  \\
		\tilde{\mathcal Q}_2(\mathbf z_k|\mathbf z^{(t)}) &=	  \left(2^\Rth-1\right) \left( \sum_{j\neq k} |\mathbf g_k^H\mathbf z_j|^2 + \sigma_0^2\right)  \nonumber \notag \\
        &- 2\mathcal Re\left\{ \left(\mathbf z_k^{(t)}\right)^H\mathbf g_k\mathbf g_k^H\mathbf z \right\} + \left\vert\mathbf g_k^H\mathbf z^{(t)}\right\vert^2,
	\end{align}
	which are all differentiable functions. According to \cite{1978A}, \textbf{Algorithm 1} converges to a Karush–Kuhn–Tucker (KKT) point if the following conditions are satisfied,
	\begin{gather}
		\mathcal Q_1(\mathbf Z) \le \tilde{\mathcal Q}_1(\mathbf Z|\mathbf Z^{(t)}), \label{ine_1} \\
		\mathcal Q_2(\mathbf z_k) \le \tilde{\mathcal Q}_2(\mathbf z_k|\mathbf z^{(t)}),\label{ine_2}\\
		\frac{\partial \mathcal Q_1(\mathbf Z)}{\partial\mathbf Z} = \frac{\partial \tilde{\mathcal Q}_1(\mathbf Z|\mathbf Z^{(t)})}{\partial\mathbf Z}, \label{deriv_1}\\
		\frac{\partial \mathcal Q_2(\mathbf z_k)}{\partial\mathbf z_k} = \frac{\partial \tilde{\mathcal Q}_2(\mathbf z_k|\mathbf z_k^{(t)})}{\partial\mathbf z_k}. \label{deriv_2}
	\end{gather}
	Obviously, (\ref{ine_1}) and (\ref{ine_2}) are satisfied. Next, we prove that (\ref{deriv_1}) and (\ref{deriv_2}) hold by taking their first order derivatives. 
	\begin{gather}
		\mathcal Q_1(\mathbf Z) = -\mathbf Z^*, \;
		\tilde{\mathcal Q}_1(\mathbf Z|\mathbf Z^{(t)}) = -\mathbf Z^{(t)*}, \\
		\frac{\partial \mathcal Q_2(\mathbf z_k)}{\partial\mathbf z_k} = \mathbf z_k^H\mathbf g_k\mathbf g_k^*,\;
		\frac{\partial \tilde{\mathcal Q}_2(\mathbf z_k|\mathbf z_k^{(t)})}{\partial\mathbf z_k} = \mathbf z_k^{(t)H}\mathbf g_k\mathbf g_k^*.
	\end{gather}
	Since $\mathbf Z = \mathbf Z^{(t)}$, and $\mathbf z_k = \mathbf z_k^{(t)}$ at stationary point, (\ref{deriv_1}) and (\ref{deriv_2}) are satisfied. Thus, \textbf{Algorithm 1} converges to a KKT point.

	\section{Maximum likelihood estimation for the angles of target}
	
	To estimate the DoAs of the target, MLE method is applied. Recalling (\ref{Rx_vec}), the received signal at the RIS sensors can be rewritten as 
	\begin{equation}
		\mathbf y_r = \alpha\mathbf p^\circ(\eta,\phi) + \mathbf n_r,
	\end{equation}
	where $\mathbf p^\circ(\eta,\phi) = \vect(\mathbf a(\eta,\phi)\mathbf b^T(\eta,\phi)\mathbf{\Theta HX})$. Assume that the RIS sensors perfectly know the BS-RIS matrix $\mathbf H$ and the transmit signal $\mathbf X$ via wired control links. Because $\mathbf n_r$ is AWGN with zeros mean and variance $\sigma_r^2\mathbf I_{M_sT}$, then with the parameters $\tilde{\bm\xi}$ provided, the likelihood function of $\mathbf y_r$ can be written as
	\begin{equation}
		f_{\mathbf y_r}(\mathbf y_r;\tilde{\bm\xi}) = \frac{1}{\sqrt{\pi\sigma_r^2M_sT}} \exp\left( -\frac{1}{\sigma_r^2} \left\Vert \mathbf y_r-\alpha\mathbf p^\circ(\eta,\phi) \right\Vert^2 \right).
	\end{equation}
	The MLE for maximizing $f_{\mathbf y_r}(\mathbf y_r;\tilde{\bm\xi})$ is equivalent to 
	\begin{equation}
		\tilde{\bm\xi} = \arg\max_{\tilde{\bm\xi}} = \arg\min_{\tilde{\bm\xi}} \left\Vert \mathbf y_r-\alpha\mathbf p^\circ(\eta,\phi) \right\Vert^2. \label{MLE}
	\end{equation}
	Hence, the complex gain $\alpha$ can be estimated as
	\begin{equation}
		\hat{\alpha} = \frac{\mathbf p^{\circ H}(\eta,\phi)}{\Vert\mathbf p^\circ(\eta,\phi)\Vert^2}\mathbf y_r. \label{alpha_hat}
	\end{equation}
	With estimated $\hat{\alpha}$, we can obtain 
	\begin{equation}
		\left\Vert \mathbf y_r-\alpha\mathbf p^\circ(\eta,\phi) \right\Vert^2 = \left\Vert \mathbf y_r\right\Vert^2 - \frac{|\mathbf p^{\circ H}(\eta,\phi)\mathbf y_r|^2}{\Vert \mathbf p^{\circ}(\eta,\phi) \Vert^2}.
	\end{equation}
	Then, $\eta$, $\phi$ can be estimated as
	\begin{equation}
		(\hat\eta, \hat\phi) = \arg\max_{\hat\eta, \hat\phi} \frac{|\mathbf p^{\circ H}(\eta,\phi)\mathbf y_r|^2}{\Vert \mathbf p^{\circ}(\eta,\phi) \Vert^2},
	\end{equation}
	where $\hat\eta$, $\hat\phi$ can be obtained by exhaustive search over $[0,\pi]$ and $[-\frac{\pi}{2},\frac{\pi}{2}]$, respectively. The MSE for estimating the azimuth and elevation angle can be computed as
	\begin{equation}
		\text{MSE} = \frac{1}{2}\mathbb E\{|\hat{\phi} - \phi|^2 + |\hat{\eta} - \eta|^2\}.
	\end{equation}
	
	\ifCLASSOPTIONcaptionsoff
	\newpage
	\fi

	\bibliographystyle{IEEEtran}
	\bibliography{Ref}
	
\end{document}